\documentclass[twocolumn]{aastex6}

\usepackage{natbib}
\usepackage{romannum}
\usepackage{longtable}
\usepackage{array} 
\usepackage{float}
\usepackage{placeins}
\newcommand{\teff}[1]{$T_\mathrm{eff}$}

\begin{document}

\title{On the Iron Abundance Anomaly in K-dwarf and Hyades Stars}

\author{Patrick D. Aleo\textsuperscript{1}, Alexander C. Sobotka\textsuperscript{1}, Ivan Ram{\'{\i}}rez\textsuperscript{2}}
\affil{\textsuperscript{1}McDonald Observatory and Department of Astronomy, University of Texas at Austin\\ 2515 Speedway, Stop C1402\\ Austin, Texas 78712-1205, USA}
\affil{\textsuperscript{2}Tacoma Community College\\ 6501 South 19th Street\\Tacoma, Washington 98466-7400, USA}

%%%%%%%%%%%%%%%%%%%%%%%%%%%%%%%%%%%%%%%%%%%%%%%%%
%%%%%%%%%%%%%%%%%%%%%%%%%%%%%%%%%%%%%%%%%%%%%%%%%

\begin{abstract}

Using standard 1D-LTE model atmosphere analysis, we provide an in-depth investigation of iron abundance as derived from neutral and singly ionization iron lines (Fe {\scriptsize{I, II}}) in nearby star clusters. Specifically, we replicate the discrepancy regarding $\Delta$[Fe/H], wherein the difference of Fe {\scriptsize{II}} - Fe {\scriptsize{I}} increases for stars of the same cluster with decreasing $T_\mathrm{eff}$, reaching an astonishing 1.0 dex at $T_\mathrm{eff}$ $\thicksim$ 4000 K. Previous studies have investigated this anomaly in the Pleiades and Hyades clusters with no concrete solution. In this analysis, we probe two samples: 63 wide binary field stars where the primary star is of sun-like temperatures and the secondary is a K-dwarf, ranging from 4231 K $\leq$ $T_\mathrm{eff}$ $\leq$ 6453 K, and 33 Hyades stars of temperatures 4268 K $\leq$ $T_\mathrm{eff}$ $\leq$ 6072 K. Previous studies have found discrepancies on the order of 1.0 dex. However, we find that these studies have neglected line-blending effects of certain Fe {\scriptsize{II}} lines, namely $\lambda$ = \{4508.29 \AA, 4993.34 \AA, 5197.58 \AA, 5325.55 \AA, 5425.26 \AA, 6456.38 \AA\}. When these lines are removed from the line-list, we find $\Delta$[Fe/H] decreases to $\thicksim$ 0.6 dex in the field binaries and $\thicksim$ 0.3 dex in the Hyades. The reason for this remaining trend is investigated by probing NLTE effects, as well as age and activity considerations using Ca {\scriptsize{II}} H+K emission and Li absorption, but these results appear to be small to negligible. 

\end{abstract}

%%%%%%%%%%%%%%%%%
%%%%%%%%%%%%%%%%%


\section{Introduction}

A long-standing problem in astrophysics stems from the work of \cite{Yong04} and \cite{schuler04:Pleiades_M34, schuler06:triplet}, who noticed startling discrepancies involving iron and oxygen abundances, leading to questions regarding the viability of standard 1D-LTE model atmospheres. \cite{Yong04} noticed that the overall iron abundance as inferred from the measurement of singly ionized iron lines rapidly increased in Hyades stars with decreasing $T_\mathrm{eff}$, starting at around 5000 K. \cite{schuler06:triplet}, likewise, found a similar trend with the inferred oxygen abundance from their measurements of the O {\scriptsize{I}} 777 nm triplet lines in dwarf stars of the Hyades, Pleiades, and M34 open clusters. \par

These findings seemingly clash with the fact that open cluster stars are formed from the same gas cloud, which inherently implies: 1) the independence of stellar atmosphere elemental abundance measurements from $T_\mathrm{eff}$, and 2) that neutral and ionized iron lines--which are a proxy for the overall iron abundance--should give identical metallicity results (within error of individual measurements). \par

First, it must be noted that there is no uniform offset in abundance for all cases as both the abundance and how quickly it increases with $T_\mathrm{eff}$ depends on the cluster. For example, in the Hyades, oxygen abundance is overestimated up to 1.0 dex at $T_\mathrm{eff} \thicksim$ 4200 K, whereas this occurs in the Pleiades at $T_\mathrm{eff} \thicksim$ 5000 K \citep{schuler06:triplet}. This is analogous to the difference between the mean iron abundance inferred from Fe {\scriptsize{I}} and Fe {\scriptsize{II}}  lines, where the Pleiades has a more rapid increase, reaching about 0.8 dex at $T_\mathrm{eff} \thicksim$ 5000 K where in the Hyades the $\Delta$[Fe/H] is around 0.6 dex overabundant at $T_\mathrm{eff} \thicksim$ 4600 K \citep{schuler10}. \par

The immediate question is whether this anomaly is due to young age, as the Hyades is $\thicksim$ 0.6 Gyr and the Pleiades is $\thicksim$ 0.1 Gyr. To date, studies from \cite{Morel:04}, \cite{Prieto04}, and \cite{ramirez07, ramirez13} have found compelling evidence that these anomalies are present in both slightly older yet chromospherically active stars as well as solar-neighborhood field dwarfs, not all of which are young. They suggest that the issue is likely more complicated, with metallicity possibly being a cofactor, and admit that our knowledge of cool dwarf atmospheres and/or spectral line formation is currently incomplete, consistent with the sentiments of \cite{schuler06:triplet, schuler10}

Currently, there are a few leading hypotheses regarding the source of the over-excitation/ionization abundance anomalies. \cite{Yong04} were one of the first to suggest non-local thermodynamic equilibrium (NLTE) effects. \cite{schuler06:triplet} subsequently employed promising ``toy models" of photospheric activity (plages, spots, and faculae) using multi-component 1D-LTE (one dimensional, static, and homogeneous atmosphere) models, suggesting magnetic activity as the potential culprit. On the other hand, work by \cite{ramirez08} has demonstrated that 1D simplifications in spectral line modeling are unlikely to be the cause. \par

These abundance anomaly results are not only troubling, but should be cause for concern from a practical point of view; iron and oxygen are crucial elements to understanding solar and Galactic chemical evolution, as pointed out by \cite{ramirez12, ramirez13}. Needless to say, our collective ability to accurately measure these and other key abundances are imperative to this process. If abundance anomalies of these magnitudes are present, significant doubt must be placed on the accuracy of these determinations in addition to our detailed knowledge of Galactic Chemical Evolution. \par

%%%%%%%%%%%%%%%%%
%%%%%%%%%%%%%%%%%

\section{Our Approach}

All previous studies regarding the oxygen and iron abundance discrepancies have involved stars in young open clusters, such as the Hyades, Pleiades, and M34, as well as the UMa moving group: systems all younger than $\thicksim$ 1 Gyr. Although it would be interesting and useful to observe cool dwarfs in older clusters such as Ruprecht 147, which is the closest known old open cluster at $\thicksim$ 3 Gyr of age and a distance of 300 pc \citep{Curtis13}, these stars are far too faint to allow us to collect high enough quality spectra with mid-sized telescopes. 

As a means to explore this topic in a different way, we decided the next best approach was to observe a number of local ($\textless$ 100 pc), bright, wide binary systems where the primary star is of Sun-like temperatures and the secondary is a K-dwarf. In this context, a binary system can be regarded as a small cluster of two stars, as they are both born from the same gas cloud. Since we are in essence modeling a cluster, more stars per system would be preferable. However, ternary and quaternary systems are far more rare, thus a two star system was optimal. As a byproduct of statistics, we would garner stars with an array of varying metallicities and ages. In this manner, we can constrain and gather these parameters independent from each other, and take advantage of their binary nature to test and ultimately identify what factor(s) lead to this apparent abundance anomaly. Observations of the Hyades itself will allow us to construct a self-consistent framework, serving as a ``control" sample for the core aspect of our study: the wide binary systems. \par

Now, with our two individual samples--those in the Hyades and those as field binary systems--we can process these spectra, measure equivalent widths (EW), and undergo data analysis in an identical manner so that we can directly compare our two samples. This way, we can both quantitatively and qualitatively explore which properties of these stars affect (if at all) the iron abundance discrepancy. 

%%%%%%%%%%%%%%%%%%%
%%%%%%%%%%%%%%%%%%%

\section{Observations \& Data Reduction}

Nearly all the spectroscopic data employed in this work were acquired using the Tull spectrograph on the 2.7\,m Harlan J.\,Smith Telescope at McDonald Observatory. With a narrow 1.2''-width slit, we were able to obtain spectra with resolution $R=\lambda/\Delta\lambda\simeq60\,000$ in the 3800 to 10\,000 \,\AA\ range, albeit with small gaps in the red portion of the spectrum. We aimed at signal-to-noise ratios of at least 200 per pixel for the brighter stars (most of the primaries) and at least 100 for the faintest ones (mainly the secondaries), leading to integration times ranging from a few minutes to one hour, depending on the stars' brightness and sky conditions. The spectra for the binary stars were taken in two runs, one in December 17--22, 2013 and one in April 17--20, 2014.

Within the same two observing runs described above we acquired very high signal-to-noise ratio ($S/N\simeq500$) spectra of a number of bright stars with effective temperatures directly measured. These stars were observed in order to calibrate the line-depth ratio versus effective temperature relations described in Section~\ref{params}.

Spectra of ``warm'' (approximately solar $T_\mathrm{eff}$) Hyades stars were taken from the high-precision abundance study by \cite{Liu16}. These data were obtained using the same configuration as that used for the binaries in this work, but in two other runs in December 2012 and December 2013. Spectra for cool Hyades were obtained during another dedicated run in December 2014 (14--16). As before, the instrumental configuration employed was identical.

Spectra for two binary pairs (HD 196755 A\&B, HD 219834 A\&B) were taken from the \cite{chaneme12} work, which is based on data taken with the MIKE spectrograph on the 6.5\,m Magellan/Clay Telescope at Las Campanas Observatory.

All spectra were reduced in the standard manner using IRAF's echelle package. After bias and overscan corrections, a pixel-to-pixel flat-field correction was applied. Spectral orders were traced using the star with the highest count on every night as reference. Spectra of ThAr lamps were used to map wavelengths on the detector; at least one ThAr exposure for every two hours was obtained to account for minor changes throughtout the night. Cosmic ray removal was done during the final extraction.

%%%%%%%%%%%%%%%%%%%
%%%%%%%%%%%%%%%%%%%

\section{Analysis \& Results}

\subsection{Spectroscopic Parameters: Binaries} \label{params}

The first step to constraining relative atmospheric parameters is to perform line strength measurements (EWs) on both the Fe {\scriptsize{I}} and Fe {\scriptsize{II}} lines in the spectra. Initially, every spectral line--which in our adopted linelist were 89 Fe {\scriptsize{I}} lines and 18 Fe {\scriptsize{II}} lines--was measured and remeasured again using IRAF's \verb|splot| task to fit Gaussian functions to the line profiles. We chose to do this by hand, without the use of automated codes, to ensure greater precision. However, not every star had a measurement for every line in the linelist. This was due to several reasons: 1) the line was too weak ($\textless$ 10 mA) and thus heavily affected by noise, 2) the spectral line fell outside the observed range, and 3) the line was too blended with other lines even after employing the \verb|deblend| technique in IRAF, a reason that will be expanded upon in later sections. \par

Once the EWs were obtained, these were transformed into the relative abundances--denoted as [Fe/H]--present in our stars' photospheres using standard 1D-LTE model atmosphere analysis. The model atmospheres were linearly interpolated within the \verb|MARCS| grid and we used MOOG \citep{Sneden73} for the spectral line calculations. We employed the Python package \verb|Qoyllur-quipu| ($q\textsuperscript{2}$)\footnote{https://github.com/astroChasqui/q2} for our stellar parameter chemical abundance analysis.\footnote{The spectra data from McDonald Observatory is available upon request.}

When running the $q\textsuperscript{2}$ code, we automatically obtain stellar parameters, namely $T_\mathrm{eff}$ and log $g$ using excitation/ionization balance. We could be fairly confident in the spectroscopic $T_\mathrm{eff}$ and log $g$ of our primary stars, as these are sunlike stars where the spectra are clean; we later used them as guides to determine the accuracy of other methods when determining measurements for these same parameters. But the K-dwarfs presented us with a problem: they are far too cool and their spectra are filled with too many line blends for accurate pure spectroscopic analysis, in addition to having more unreliable model atmospheres. \par

To alleviate the uncertainties for secondaries, we used isochrones. Since our binaries are born of the same gas cloud, they should, in theory, be on the same isochrone. So, after determining $T_\mathrm{eff}$ and log $g$ of our primaries, we needed reliable $T_\mathrm{eff}$ of our secondaries from which we could use our isochrone data to interpolate their log $g$. \par

To do this, it was imperative for us to be as precise and accurate as possible with our $T_\mathrm{eff}$ determinations, as any error would propagate into log $g$. Further, we needed to be precise and consistent across our binary and Hyades samples for determining $T_\mathrm{eff}$, as well as not employ spectroscopic measurements directly where possible. In order to determine effective temperatures in a nearly model independent fashion, we calibrated line-depth ratio (LDR) versus $T_\mathrm{eff}$ relations using a sample of stars for which effective temperatures have been measured directly. The sample of calibrators was taken from the works by \cite{boyajian12, boyajian13}, who measured angular diameters using interferometry to determine direct $T_\mathrm{eff}$ values. A total of 35 stars were used for this calibration.

\begin{figure}[H]
\centering
\includegraphics[width=1\linewidth]{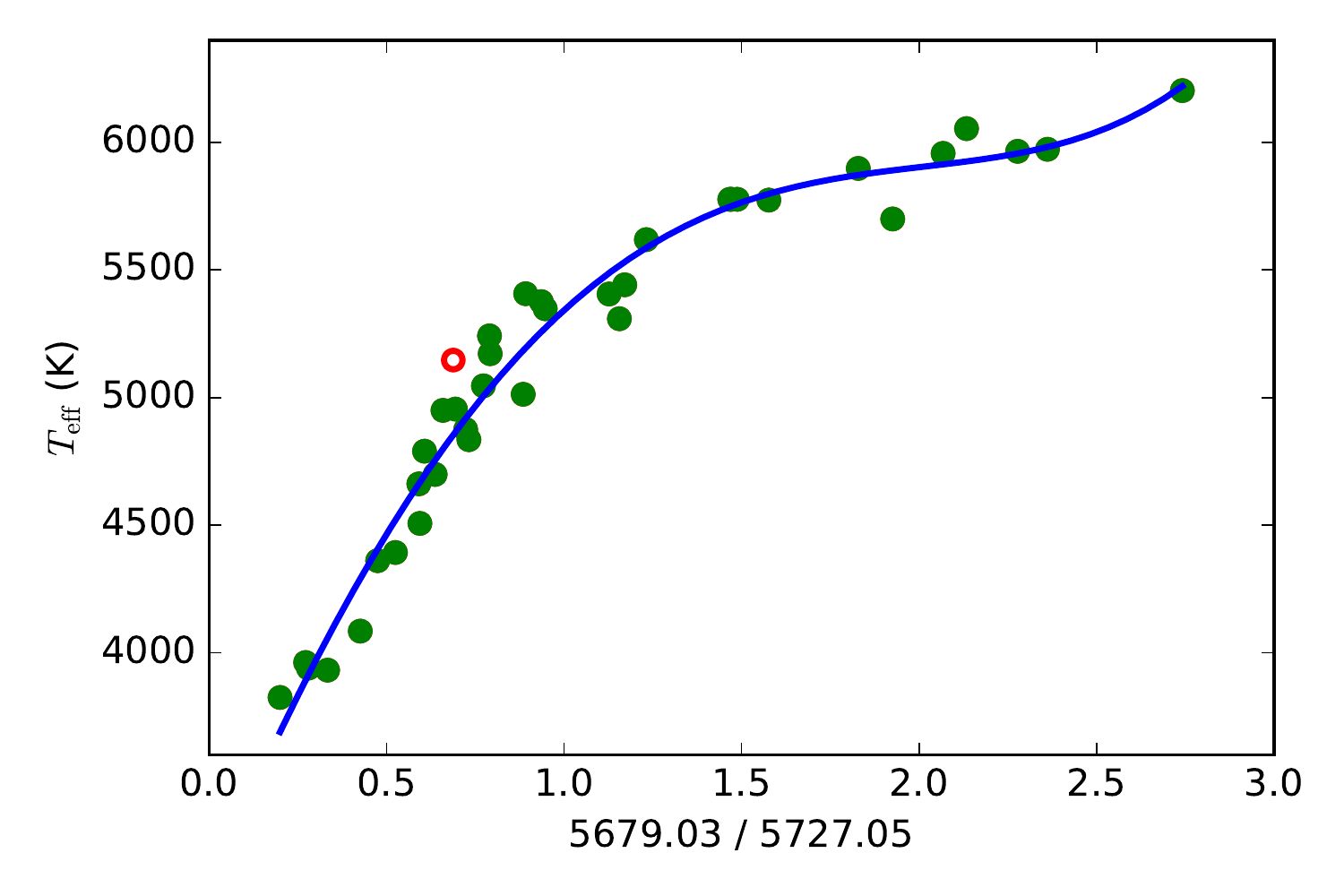}
\vspace{-0.3cm}
\caption{An example of the LDR-$T_\mathrm{eff}$ relation derived for the pair $\lambda_{1}$ = 5679.03 {\AA } (Fe {\scriptsize{I}}, $\chi$ = 4.65, log $gf$ = -0.756), $\lambda_{2}$ = 5727.05 {\AA } (V {\scriptsize{I}}, $\chi$ = 1.08, log $gf$ = 0). The green dots are stars, and the red circular shell is an outlier, excluded in the calculation.}
\label{LDR}
\end{figure}

We examined our spectra to find the best spectral line pairs for effective temperature determination using the LDR technique, described, for example, in \cite{Grey91}. We searched for pairs from the list by \cite{Sousa10} that would give us the tightest LDR vs $T_\mathrm{eff}$ correlation. The 10 best line pairs were finally adopted. An example of the LDR-$T_\mathrm{eff}$ relation derived for one of those pairs is shown in Figure \ref{LDR}. On average, the standard deviation of each of these 10 LDR-$T_\mathrm{eff}$ relations was just below 100\,K, with standard error $\sim$35\,K. A 1-$\sigma$ clipping criterion was adopted in the calibrations to remove outliers (between zero and ten).

When these calibrations were applied to our solar spectra, we obtained an average of $5745\pm8$\,K, where the error bar was standard error. To account for this offset (the Sun's nominal effective temperature is 5777\,K) we shifted all temperatures derived using these calibrations up by 32\,K. Interestingly, the LDR calibrations provide more accurate temperatures for the K-dwarfs than the sun-like stars. On average, the standard deviation of the $T_\mathrm{eff, LDR}$ - $T_\mathrm{eff, Direct}$ values is about 100\,K at solar $T_\mathrm{eff}$, but less than 60\,K for stars with $T_\mathrm{eff}$ below 4750\,K.

We applied these calibrations to both the binary sample as well as the Hyades sample to assign a $T_\mathrm{eff}$ value to each star. The standard deviation from the 10 calibrations was adopted as the $T_\mathrm{eff}$ error. We also compared the results from the spectroscopically-derived temperatures and the line-depth temperatures in our primary stars to see if they yielded similar results. On average they were in good agreement with measurements within 40 K. \par

Our goal is to ultimately obtain a self-consistent set of $T_\mathrm{eff}$, log $g$, [Fe/H], and $\xi$ (microturbulent velocity) for both our binary stars sample and our Hyades sample. Thus, it is important to use the LDR temperatures for all stars in both samples where possible. There are a few stars for which we could not adopt LDR, because the difference in the LDR-derived temperature and the spectroscopic temperatures was around 300 K; these are the hottest stars in our sample, with $T_\mathrm{eff}$ $\textgreater$ 6300 K, and the LDR method could not compensate for these extremely high values. Therefore, we used the spectroscopic temperatures obtained with $q\textsuperscript{2}$ for these select stars (indicated by * in tables). We were confident in doing this as our LDR and spectroscopic temperatures were in good agreement for all other temperature values, in particular after applying the temperature shift. Now with $T_\mathrm{eff}$ determined using LDR, we calculated new log $g$ and [Fe/H] (we calculated $\xi$ later) with their errors for the primaries, and then used the recently obtained LDR temperatures of all our stars combined with isochrone data from the Dartmouth Stellar Evolution Database to interpolate in log $g$ for our secondaries. \par

With our first round of calculations, we used parameters that were dependent upon previously determined values. But we needed an independent set for both our binary and Hyades sample to ensure that we could test different parameters individually and determine what factor(s) may result in the abundance anomaly. The Python package $q\textsuperscript{2}$ is able to calculate log $g$ using derived parameters including parallax, visual magnitude (V\textsubscript{mag}), and their errors. We obtained parallax from Hipparcos and V\textsubscript{mag} from The General Catalogue of Photometric Data (GCPD), as provided by \cite{GCPD}.\footnote{http://obswww.unige.ch/gcpd/gcpd.html} Since many of the parallax data were not available for the secondaries, we adopted the value from its sunlike companion. The only stars without a parallax were HD 34254a and HD 34254b, and the only star without a V\textsubscript{mag} value was HD 200660b, so we adopted their spectroscopic log $g$ for both HD 34254a and HD 34254b as well as HD 200660a and HD 200660b. Despite HD 200660a having a parallax value, we wanted to be consistent with the type of measurement within a binary, and we felt confident in this approach as the spectroscopic and $q\textsuperscript{2}$-derived measurements were in good agreement when comparing all log $g$ data across our sample. 

Finally, we performed one more total iteration for every parameter, namely $T_\mathrm{eff}$, log $g$, [Fe/H], and $\xi$ using $q\textsuperscript{2}$. We calculated the microturbulent velocity after adopting the empirical formula from \cite{ramirez13}, which predicts $\xi$-values to a 1-$\sigma$ error of .12 km s\textsuperscript{-1}. The reason as to why this parameter was calculated last is that it requires the input of  $T_\mathrm{eff}$, log $g$, and [Fe/H], so we needed to make sure these values were finalized. Now we had an independent set: precise and accurate $T_\mathrm{eff}$ from LDR using ``direct" calibration; log $g$ from isochrones, parallax and V\textsubscript{mag}; [Fe/H] as well as Fe {\scriptsize{I}} and Fe {\scriptsize{II}} from spectroscopic measurements; $\xi$ from an empirically derived formula.

Our final stellar parameters for our field binaries are presented in Table \ref{binarystars}, after line-blended lines have been removed (See Section \ref{lb}). 

\LTcapwidth=\textwidth
\begin{longtable*}{lllrrrrrrrrrrr}
\caption{\label{binarystars} Stellar Parameters of Binary Stars}\\
\hline\hline
HD & $T_\mathrm{eff}$  & err $T_\mathrm{eff}$ & log $g$ & err log $g$ & [Fe/H] & err [Fe/H] & Fe {\scriptsize{I}} & err Fe {\scriptsize{I}} & Fe {\scriptsize{II}} & err Fe {\scriptsize{II}} &$\xi$ & nFe {\scriptsize{I}} & nFe {\scriptsize{II}} \\ 
\hline
\endfirsthead
\caption{continued.}\\
\hline\hline
HD & $T_\mathrm{eff}$  & err $T_\mathrm{eff}$ & log $g$ & err log $g$ & [Fe/H] & err [Fe/H] & Fe {\scriptsize{I}} & err Fe {\scriptsize{I}} & Fe {\scriptsize{II}} & err Fe {\scriptsize{II}} &$\xi$ & nFe {\scriptsize{I}} & nFe {\scriptsize{II}} \\ 
\hline
\endhead
\hline
\endfoot
%HD       Teff     erT    logg    err log    feh        err fe      feh1    err fe1    fe2       err fe II    mic    nfe1  nfe2
2567a & 5814 & 25 & 4.16 & 0.041 & 0.065 & 0.068 & 0.054 & 0.062	 & 0.146 & 0.056 & 1.24 & 87 & 12 \\
2567b & 5081 & 48 & 4.45 & 0.037 & 0.117 & 0.074 & 0.105 & 0.063 & 0.248 & 0.079 & 0.52 & 88 & 8 \\
4614a & 6032 & 46 & 4.43 & 0.029 & -0.230 & 0.047 & -0.220 & 0.039 & -0.303 & 0.044 & 1.29 & 74 & 10 \\
4614b$^{\phi}$ & 4011 & 38 & 4.75 & 0.005 & -0.305 & 0.448 & -0.383 & 0.213 & N/A & N/A & 0.00 & 76 & 0 \\
7439a* & 6461 & 40 & 4.10 & 0.036 & -0.306 & 0.060 & -0.305 & 0.058 & -0.311 & 0.078 & 1.79 & 64 & 10 \\
7439b & 5282 & 46 & 4.61 & 0.023 & -0.278 & 0.044 & -0.279 & 0.044 & -0.265 & 0.054 & 0.62 & 87 & 8 \\
8009a & 5819 & 62 & 4.35 & 0.089 & -0.202 & 0.057 & -0.190 & 0.040 & -0.294 & 0.085 & 1.16 & 82 & 11 \\
8009b & 5296 & 47 & 4.49 & 0.055 & -0.217 & 0.046 & -0.218 & 0.044 & -0.200 & 0.069 & 0.69 & 85 & 8 \\
13043a & 5818 & 17 & 4.16 & 0.029 & -0.023 & 0.048 & -0.026 & 0.048 & 0.000 & 0.049 & 1.25 & 86 & 11 \\
13043b & 4375 & 35 & 4.70 & 0.017 & -0.217 & 0.164 & -0.218 & 0.163 & -0.195 & 0.264 & 0.00 & 84 & 3 \\
13357a & 5729 & 44 & 4.48 & 0.043 & -0.063 & 0.050 & -0.058 & 0.045 & -0.099 & 0.077 & 1.02 & 88 & 11 \\
13357b & 5436 & 45 & 4.54 & 0.044 & -0.053 & 0.047 & -0.052 & 0.041 & -0.066 & 0.086 & 0.76 & 89 & 11 \\
33334a & 5660 & 58 & 4.41 & 0.063 & -0.021 & 0.045 & -0.025 & 0.043 & 0.015 & 0.052 & 1.00 & 87 & 10 \\
33334b & 5221 & 65 & 4.51 & 0.044 & -0.064 & 0.060 & -0.071 & 0.056 & 0.017 & 0.046 & 0.61 & 89 & 7 \\
34254a & 5773 & 78 & 4.09 & 0.056 & -0.311 & 0.062 & -0.307 & 0.062 & -0.343 & 0.057 & 1.26 & 81 & 11 \\
34254b & 5371 & 129 & 4.52 & 0.076 & -0.176 & 0.065 & -0.181 & 0.064 & -0.121 & 0.050 & 0.73 & 87 & 8 \\
35961a & 5909 & 39 & 4.38 & 0.045 & -0.207 & 0.053 & -0.199 & 0.043 & -0.268 & 0.080 & 1.22 & 81 & 11 \\
35961b & 4744 & 62 & 4.63 & 0.026 & -0.251 & 0.092 & -0.259 & 0.088 & -0.138 & 0.098 & 0.19 & 85 & 6 \\
38393a* & 6339 & 23 & 4.31 & 0.029 & -0.104 & 0.064 & -0.099 & 0.062 & -0.133 & 0.075 & 1.58 & 73 & 12 \\
38393b & 4856 & 36 & 4.57 & 0.020 & -0.069 & 0.085 & -0.083 & 0.068 & 0.089 & 0.099 & 0.30 & 88 & 8 \\
69056a & 5505 & 51 & 4.28 & 0.038 & 0.024 & 0.048 & 0.021 & 0.042 & 0.044 & 0.078 & 0.94 & 87 & 12 \\
69056b & 4663 & 46 & 4.63 & 0.024 & -0.075 & 0.107 & -0.079 & 0.098 & -0.007 & 0.217 & 0.12 & 85 & 5 \\
73668a & 5931 & 66 & 4.37 & 0.046 & -0.073 & 0.040 & -0.072 & 0.040 & -0.081 & 0.042 & 1.23 & 83 & 10 \\
73668b & 5276 & 54 & 4.49 & 0.036 & -0.026 & 0.049 & -0.026 & 0.049 & -0.020 & 0.053 & 0.66 & 85 & 5 \\
78154a* & 6441 & 28 & 4.17 & 0.029 & 0.009 & 0.067 & 0.022 & 0.055 & -0.072 & 0.083 & 1.73 & 77 & 12 \\
78154b & 4551 & 55 & 4.60 & 0.021 & -0.225 & 0.146 & -0.236 & 0.126 & -0.053 & 0.297 & 0.05 & 89 & 6 \\
90839a & 6163 & 28 & 4.42 & 0.024 & -0.128 & 0.038 & -0.127 & 0.038 & -0.139 & 0.041 & 1.39 & 77 & 11 \\
90839b & 4231 & 42 & 4.73 & 0.016 & -0.338 & 0.223 & -0.345 & 0.217 & 0.239 & N/A & 0.00 & 84 & 1 \\
94979a & 5465 & 47 & 4.47 & 0.061 & -0.041 & 0.038 & -0.045 & 0.035 & -0.009 & 0.048 & 0.82 & 87 & 11 \\
94979b & 4544 & 63 & 4.66 & 0.029 & -0.049 & 0.111 & -0.054 & 0.109 & 0.118 & 0.025 & 0.01 & 87 & 3 \\
99491a & 5281 & 67 & 4.38 & 0.036 & 0.299 & 0.089 & 0.271 & 0.052 & 0.497 & 0.046 & 0.70 & 87 & 12 \\
99491b & 4814 & 46 & 4.53 & 0.026 & 0.341 & 0.105 & 0.325 & 0.091 & 0.529 & 0.064 & 0.26 & 89 & 8 \\
108361a & 5661 & 87 & 4.22 & 0.076 & -0.070 & 0.041 & -0.070 & 0.039 & -0.072 & 0.057 & 1.10 & 86 & 12 \\
108361b & 4923 & 52 & 4.60 & 0.032 & -0.088 & 0.095 & -0.105 & 0.071 & 0.117 & 0.114 & 0.33 & 83 & 7 \\
114146a & 5681 & 51 & 4.15 & 0.074 & 0.193 & 0.065 & 0.177 & 0.054 & 0.296 & 0.028 & 1.13 & 81 & 12 \\
114146b & 4789 & 65 & 4.50 & 0.038 & 0.431 & 0.151 & 0.424 & 0.145 & 0.518 & 0.230 & 0.25 & 77 & 6 \\
118576a & 5793 & 27 & 4.42 & 0.145 & -0.117 & 0.052 & -0.113 & 0.053 & -0.144 & 0.034 & 1.10 & 88 & 12 \\
118576b & 5041 & 83 & 4.58 & 0.087 & -0.089 & 0.099 & -0.116 & 0.060 & 0.126 & 0.089 & 0.43 & 87 & 11 \\
120066a & 5742 & 39 & 4.06 & 0.032 & 0.020 & 0.047 & 0.011 & 0.042 & 0.087 & 0.018 & 1.24 & 85 & 12 \\
120066b & 4365 & 55 & 4.67 & 0.019 & -0.051 & 0.119 & -0.056 & 0.114 & 0.054 & 0.200 & 0.00 & 84 & 4 \\
123453a & 6065 & 79 & 4.14 & 0.075 & 0.032 & 0.058 & 0.026 & 0.057 & 0.075 & 0.050 & 1.45 & 85 & 12 \\
123453b & 5650 & 45 & 4.46 & 0.061 & 0.016 & 0.057 & 0.009 & 0.055 & 0.069 & 0.041 & 0.97 & 85 & 11 \\
131023a & 5516 & 37 & 4.42 & 0.037 & 0.184 & 0.060 & 0.169 & 0.046 & 0.292 & 0.044 & 0.87 & 89 & 12 \\
131023b & 4494 & 46 & 4.67 & 0.020 & 0.189 & 0.140 & 0.176 & 0.124 & 0.404 & 0.233 & 0.00 & 84 & 5 \\
131156a & 5590 & 42 & 4.56 & 0.021 & -0.007 & 0.074 & -0.001 & 0.075 & -0.059 & 0.046 & 0.87 & 88 & 10 \\
131156b & 4359 & 38 & 4.69 & 0.016 & -0.082 & 0.152 & -0.094 & 0.134 & 0.125 & 0.292 & 0.00 & 89 & 5 \\
138004a & 5799 & 66 & 4.48 & 0.033 & -0.098 & 0.027 & -0.096 & 0.028 & -0.108 & 0.019 & 1.08 & 85 & 12 \\
138004b & 4327 & 72 & 4.63 & 0.020 & -0.072 & 0.185 & -0.092 & 0.164 & 0.279 & 0.218 & 0.00 & 88 & 5 \\
151090a & 5041 & 105 & 3.56 & 0.087 & -0.224 & 0.040 & -0.225 & 0.041 & -0.216 & 0.031 & 0.95 & 86 & 12 \\
151090b & 4461 & 62 & 4.61 & 0.022 & -0.259 & 0.114 & -0.270 & 0.103 & -0.033 & 0.143 & 0.00 & 89 & 4 \\
158226a & 5895 & 123 & 4.27 & 0.079 & -0.402 & 0.062 & -0.392 & 0.057 & -0.479 & 0.050 & 1.27 & 84 & 11 \\
158226b & 5370 & 176 & 4.51 & 0.046 & -0.421 & 0.065 & -0.430 & 0.061 & -0.338 & 0.035 & 0.74 & 86 & 9 \\
187013a* & 6453 & 36 & 4.21 & 0.024 & 0.002 & 0.056 & -0.001 & 0.057 & 0.020 & 0.045 & 1.72 & 76 & 12 \\
187013b & 4639 & 33 & 4.62 & 0.019 & -0.090 & 0.115 & -0.109 & 0.089 & 0.147 & 0.153 & 0.10 & 87 & 7 \\
195019a & 5759 & 43 & 4.13 & 0.042 & 0.040 & 0.033 & 0.037 & 0.033 & 0.067 & 0.030 & 1.21 & 87 & 12 \\
195019b & 4310 & 22 & 4.64 & 0.014 & -0.020 & 0.234 & -0.044 & 0.211 & 0.488 & 0.111 & 0.00 & 85 & 4 \\
196755a & 5706 & 65 & 3.71 & 0.023 & 0.043 & 0.027 & 0.046 & 0.027 & 0.018 & 0.021 & 1.38 & 80 & 9 \\
196755b & 4929 & 36 & 4.53 & 0.022 & 0.026 & 0.060 & 0.017 & 0.044 & 0.155 & 0.098 & 0.37 & 80 & 6 \\
200660a & 5563 & 21 & 4.44 & 0.059 & 0.082 & 0.071 & 0.080 & 0.075 & 0.098 & 0.037 & 0.90 & 83 & 12 \\
200660b & 5049 & 42 & 4.57 & 0.019 & 0.204 & 0.092 & 0.194 & 0.087 & 0.349 & 0.059 & 0.43 & 88 & 6 \\
213013a & 5359 & 61 & 4.45 & 0.053 & 0.050 & 0.042 & 0.046 & 0.040 & 0.089 & 0.045 & 0.74 & 85 & 10 \\
213013b & 4980 & 31 & 4.53 & 0.039 & 0.083 & 0.061 & 0.076 & 0.056 & 0.154 & 0.070 & 0.40 & 87 & 9 \\
219834a & 5422 & 80 & 3.87 & 0.042 & 0.180 & 0.043 & 0.178 & 0.044 & 0.203 & 0.025 & 1.07 & 81 & 9 \\
219834b & 4999 & 51 & 4.39 & 0.036 & 0.195 & 0.058 & 0.187 & 0.051 & 0.284 & 0.065 & 0.48 & 80 & 7 \\
vesta & 5777 & 0 & 4.44 & 0.000 & 0.000 & 0.000 & 0.000 & 0.000 & 0.000 & 0.000 & 1.03 & 0 & 0 \\ [1ex] 
\hline
\caption{List of final stellar parameters for each primary and secondary star. Here, nFe {\scriptsize{I}} is the number of Fe {\scriptsize{I}} lines from the linelist used when determining abundance. Likewise, nFe {\scriptsize{II}} is the number of Fe {\scriptsize{II}} lines used to determine abundance. Again, * is used to identify for which stars we adopted spectroscopic $T_\mathrm{eff}$ instead of temperatures determined by LDR. The symbol $\phi$ denotes the star HD 4614b, which had zero accurate Fe {\scriptsize{II}} measurements due to line-blends.}
\end{longtable*}

%%%%%%%%%%%%%%%%%%%%%

\subsection{Spectroscopic Parameters: Hyades}

In keeping with our approach, we determined the stellar parameters for our Hyades sample in a similar manner to that of our binaries. Specifically, we developed an independent set of $T_\mathrm{eff}$, log $g$, [Fe/H], and $\xi$ using $q\textsuperscript{2}$. However, there is one subtle difference. Since we had parallax measurements for all of our Hyades sample stars except HD27835, in which case we adopted spectroscopic log $g$, we used their values in addition to isochrones to infer log $g$. This provided us with a more precise estimate for log $g$, as using isochrones alone relies on the assumption that all stars in the cluster are of exactly the same age. Also, in regard to temperature, LDR provided all temperatures values for our Hyades sample, as none had $T_\mathrm{eff}$ $\textgreater$ 6300 K. Now, similar to that of the binaries, we had a full independent set of parameters with which we could use to 1) replicate the anomaly similar to what \cite{Yong04} found in the Hyades, and 2) compare our results with binaries. Our final Hyades parameters are presented in Table \ref{Hyadesstars}, after line-blended lines have been removed (See Section \ref{lb}).  

\raggedbottom
\begin{longtable*}{lllrrrrrrrrrrr}
\caption{\label{Hyadesstars} Stellar Parameters of Hyades Stars}\\
\hline\hline
id & $T_\mathrm{eff}$  & err $T_\mathrm{eff}$ & log $g$ & err log $g$ & [Fe/H] & err [Fe/H] & Fe {\scriptsize{I}} & err Fe {\scriptsize{I}} & Fe {\scriptsize{II}} & err Fe {\scriptsize{II}} &$\xi$ & nFe {\scriptsize{I}} & nFe {\scriptsize{II}} \\ 
\hline
\endfirsthead
\caption{continued.}\\
\hline\hline
id & $T_\mathrm{eff}$  & err $T_\mathrm{eff}$ & log $g$ & err log $g$ & [Fe/H] & err [Fe/H] & Fe {\scriptsize{I}} & err Fe {\scriptsize{I}} & Fe {\scriptsize{II}} & err Fe {\scriptsize{II}} &$\xi$ & nFe {\scriptsize{I}} & nFe {\scriptsize{II}} \\ 
\hline
\endhead
\hline
\endfoot
%HD       Teff     erT    logg    err log    feh        err fe      feh1    err fe1    fe2       err fe II    mic    nfe1  nfe2
HD27835$^{\S}$ & 5862 & 26 & 4.40 & 0.042 & 0.082 & 0.063 & 0.066 & 0.049 & 0.195 & 0.034 & 1.15 & 87 & 12 \\
HIP14976 & 5505 & 68 & 4.45 & 0.044 & 0.160 & 0.050 & 0.151 & 0.042 & 0.227 & 0.054 & 0.85 & 87 & 12 \\
HIP15563 & 4553 & 50 & 4.64 & 0.019 & 0.061 & 0.127 & 0.045 & 0.100 & 0.263 & 0.244 & 0.01 & 76 & 6 \\
HIP17766 & 4268 & 59 & 4.71 & 0.014 & -0.096 & 0.190 & -0.107 & 0.173 & 0.182 & 0.424 & 0.00 & 75 & 3 \\
HIP18018 & 4537 & 40 & 4.67 & 0.019 & 0.068 & 0.129 & 0.052 & 0.108 & 0.278 & 0.205 & 0.00 & 77 & 6 \\
HIP18322 & 4705 & 53 & 4.64 & 0.022 & 0.167 & 0.141 & 0.152 & 0.120 & 0.269 & 0.228 & 0.13 & 87 & 12 \\
HIP18946 & 4721 & 56 & 4.63 & 0.026 & 0.182 & 0.127 & 0.169 & 0.116 & 0.276 & 0.168 & 0.15 & 87 & 12 \\
HIP19148 & 5907 & 15 & 4.41 & 0.041 & 0.076 & 0.065 & 0.064 & 0.058 & 0.162 & 0.049 & 1.18 & 87 & 12 \\
HIP19263 & 4803 & 48 & 4.57 & 0.029 & 0.159 & 0.116 & 0.144 & 0.101 & 0.264 & 0.167 & 0.24 & 87 & 12 \\
HIP19316 & 4301 & 54 & 4.71 & 0.014 & -0.090 & 0.219 & -0.102 & 0.159 & 0.137 & 0.750 & 0.00 & 75 & 4 \\
HIP19441 & 4504 & 43 & 4.68 & 0.017 & 0.011 & 0.129 & 0.001 & 0.117 & 0.137 & 0.216 & 0.00 & 76 & 6 \\
HIP19781 & 5655 & 69 & 4.50 & 0.039 & 0.158 & 0.040 & 0.152 & 0.037 & 0.201 & 0.038 & 0.94 & 87 & 12 \\
HIP19786 & 5769 & 62 & 4.46 & 0.039 & 0.146 & 0.055 & 0.135 & 0.046 & 0.229 & 0.045 & 1.05 & 87 & 12 \\
HIP19793 & 5751 & 60 & 4.41 & 0.047 & 0.145 & 0.052 & 0.137 & 0.049 & 0.201 & 0.040 & 1.07 & 87 & 12 \\
HIP20082 & 4882 & 64 & 4.54 & 0.034 & 0.202 & 0.104 & 0.189 & 0.096 & 0.299 & 0.116 & 0.32 & 87 & 12 \\
HIP20146 & 5574 & 76 & 4.45 & 0.046 & 0.205 & 0.049 & 0.201 & 0.049 & 0.229 & 0.042 & 0.90 & 87 & 12 \\
HIP20237 & 6003 & 68 & 4.35 & 0.035 & 0.086 & 0.089 & 0.070 & 0.078 & 0.201 & 0.087 & 1.29 & 87 & 12 \\
HIP20485 & 4465 & 56 & 4.66 & 0.020 & 0.048 & 0.167 & 0.019 & 0.118 & 0.364 & 0.291 & 0.00 & 76 & 7 \\
HIP20492 & 5203 & 56 & 4.53 & 0.045 & 0.247 & 0.076 & 0.240 & 0.074 & 0.296 & 0.079 & 0.58 & 87 & 12 \\
HIP20577 & 5894 & 45 & 4.37 & 0.047 & 0.069 & 0.058 & 0.063 & 0.056 & 0.119 & 0.052 & 1.20 & 87 & 12 \\
HIP20741 & 5725 & 46 & 4.42 & 0.048 & 0.155 & 0.044 & 0.148 & 0.039 & 0.210 & 0.038 & 1.04 & 87 & 12 \\
HIP20815 & 6049 & 57 & 4.30 & 0.041 & 0.056 & 0.098 & 0.042 & 0.092 & 0.162 & 0.081 & 1.35 & 87 & 12 \\
HIP20826 & 6072 & 52 & 4.35 & 0.036 & 0.068 & 0.078 & 0.061 & 0.075 & 0.121 & 0.082 & 1.34 & 87 & 12 \\
HIP20899 & 5875 & 48 & 4.39 & 0.039 & 0.101 & 0.070 & 0.090 & 0.064 & 0.181 & 0.067 & 1.18 & 87 & 12 \\
HIP20949 & 5444 & 62 & 4.45 & 0.059 & 0.203 & 0.056 & 0.200 & 0.052 & 0.219 & 0.079 & 0.80 & 87 & 12 \\
HIP21112 & 6039 & 48 & 4.33 & 0.037 & 0.027 & 0.063 & 0.015 & 0.056 & 0.116 & 0.034 & 1.33 & 87 & 12 \\
HIP21138 & 4330 & 25 & 4.69 & 0.015 & -0.014 & 0.192 & -0.032 & 0.152 & 0.260 & 0.456 & 0.00 & 75 & 5 \\
HIP21317 & 5839 & 48 & 4.41 & 0.041 & 0.114 & 0.056 & 0.104 & 0.049 & 0.189 & 0.046 & 1.14 & 87 & 12 \\
HIP21543 & 5888 & 44 & 4.20 & 0.066 & 0.100 & 0.075 & 0.098 & 0.078 & 0.109 & 0.057 & 1.27 & 87 & 12 \\
HIP21637 & 5924 & 34 & 4.33 & 0.033 & 0.042 & 0.065 & 0.029 & 0.059 & 0.132 & 0.033 & 1.24 & 87 & 12 \\
HIP22177 & 4401 & 48 & 4.68 & 0.017 & 0.001 & 0.168 & -0.017 & 0.136 & 0.273 & 0.350 & 0.00 & 75 & 5 \\
HIP22422 & 5918 & 33 & 4.33 & 0.035 & 0.076 & 0.070 & 0.062 & 0.062 & 0.173 & 0.044 & 1.23 & 87 & 12 \\
HIP23312 & 4979 & 51 & 4.54 & 0.037 & 0.231 & 0.087 & 0.218 & 0.075 & 0.333 & 0.115 & 0.39 & 87 & 11 \\
\hline
\caption{List of final stellar parameters for each Hyades star. The symbol {\S} denotes that there was no parallax for this star, so we used spectroscopic log $g$ and assumed err log $g$ using the average from similar temperature stars.}
\end{longtable*}

%%%%%%%%%%%%%%%%%%%%%

\subsection{The Line-Blending Problem} \label{lb}

As is well known, cooler stars have complex spectra: the absorption lines, in general, become stronger and occasionally blend together to artificially increase the abundance in an element. When EW measurements are taken, it is possible to measure the combined EW of two or more elements at a specific wavelength instead of a single line. This is known as \textbf{line-blending}, and it can quickly become problematic when calculating stellar abundances. In cooler stars, spectral lines blend together and become both wider and have noticeable bumps in what should be a clean Gaussian profile. Take the case when we are measuring a specific Fe {\scriptsize{II}} line: if this is blended with a Ca {\scriptsize{I}} line, we are in fact measuring the combined EW of the desired Fe {\scriptsize{II}} line and an undesired Ca {\scriptsize{I}} line. This leads to false over-excitation/ionization abundances, and when unchecked, leads to false and overabundant measurements. Not surprisingly, line blending was a reasonable suggestion for the main cause of the current anomaly we observe. 

Across the board, our stars are far more abundant in neutral Fe {\scriptsize{I}} lines than the singly ionized Fe {\scriptsize{II}} lines. As we go cooler in $T_\mathrm{eff}$, some of the Fe {\scriptsize{II}} lines become nearly impossible to accurately measure due to line-blends creating unacceptably high abundances or weak lines becoming non-differentiable from noise. In our analysis of our original 18 Fe {\scriptsize{II}} lines, 6 consistently yielded very high abundances, on the order of a full magnitude greater than what is inferred from the Fe {\scriptsize{I}} lines. In Tables \ref{binarystars} and \ref{Hyadesstars}, it can be seen that no star has more than 12 measured Fe {\scriptsize{II}} lines; when investigating the abundances as inferred from Fe {\scriptsize{II}} lines, these six lines were consistently over abundant for multiple stars. In addition, the Gaussian profiles for these Fe {\scriptsize{II}} lines were no longer centered about the expected wavelength. The combination of these factors lead us to believe that these lines were most likely line-blended. So, in $both$ our binary star sample and our Hyades sample we removed the following lines: $\lambda$ = \{4508.29 \AA, 4993.34 \AA, 5197.58 \AA, 5325.55 \AA, 5425.26 \AA, 6456.38 \AA\}. \par

As a visual representation, let us look at spectra from both a ``good" Fe {\scriptsize{II}} line and a ``bad" Fe {\scriptsize{II}} line. Here, we consider two binary pairs: HD131156a ($T_\mathrm{eff}$ = 5590 K), HD131156b ($T_\mathrm{eff}$ = 4359 K); HD138004a ($T_\mathrm{eff}$ = 5799 K), HD138004b ($T_\mathrm{eff}$ = 4327 K). \par

\begin{figure}[!t]
  \centering
  \begin{minipage}[b]{0.45\textwidth}
    \includegraphics[width=\textwidth]{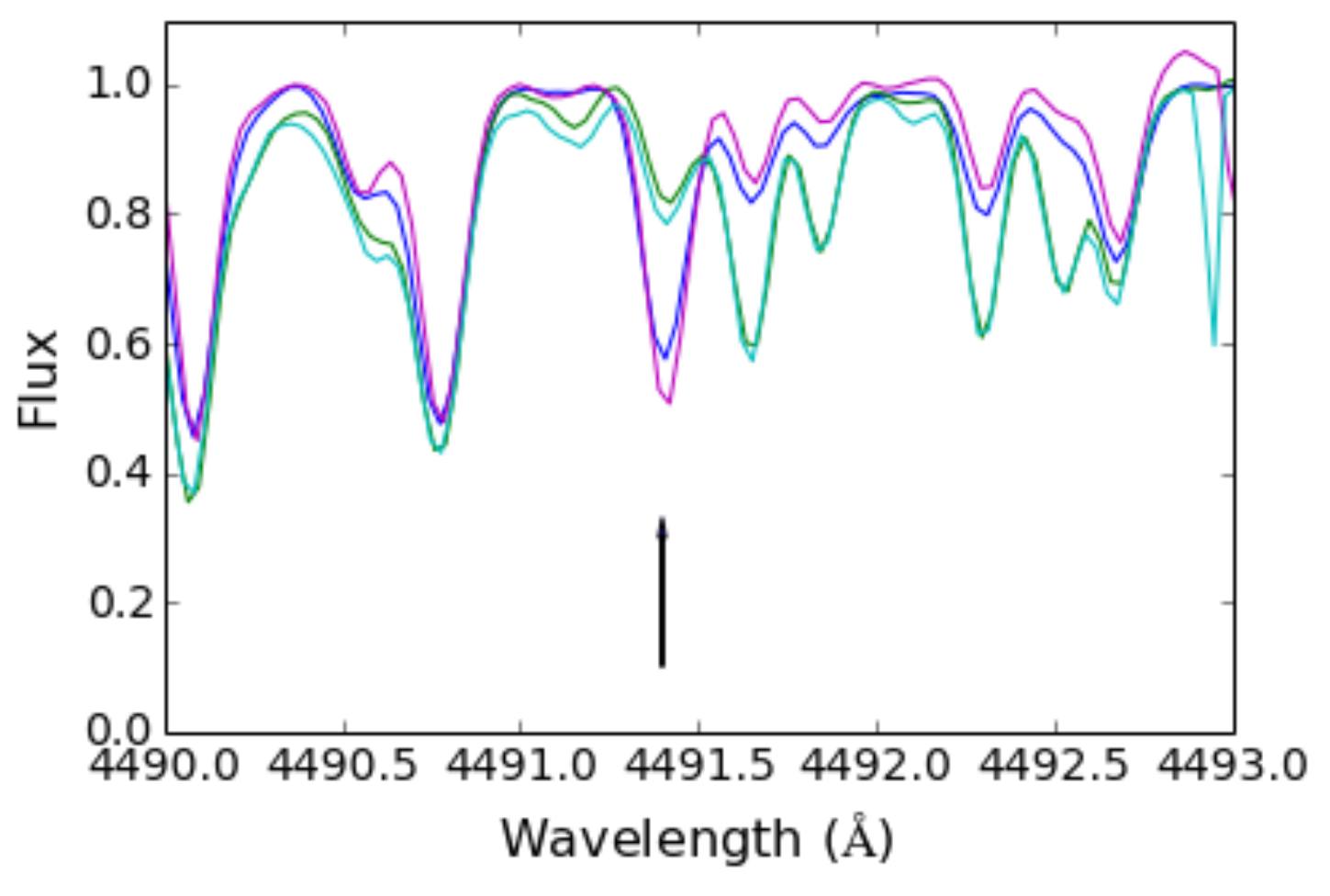}
    \caption{The spectra of binary stars HD131156a (dark blue), HD131156b (green), HD138004a (magenta), and HD138004b (cyan), with the arrow pointing at the region of interest: the ``good" Fe {\scriptsize{II}} line at $\lambda$ = 4491.4 \AA.}
    \label{BinGoodFeII}
  \end{minipage}%
  \hfill
    \begin{minipage}[b]{0.45\textwidth}
    \includegraphics[width=\textwidth]{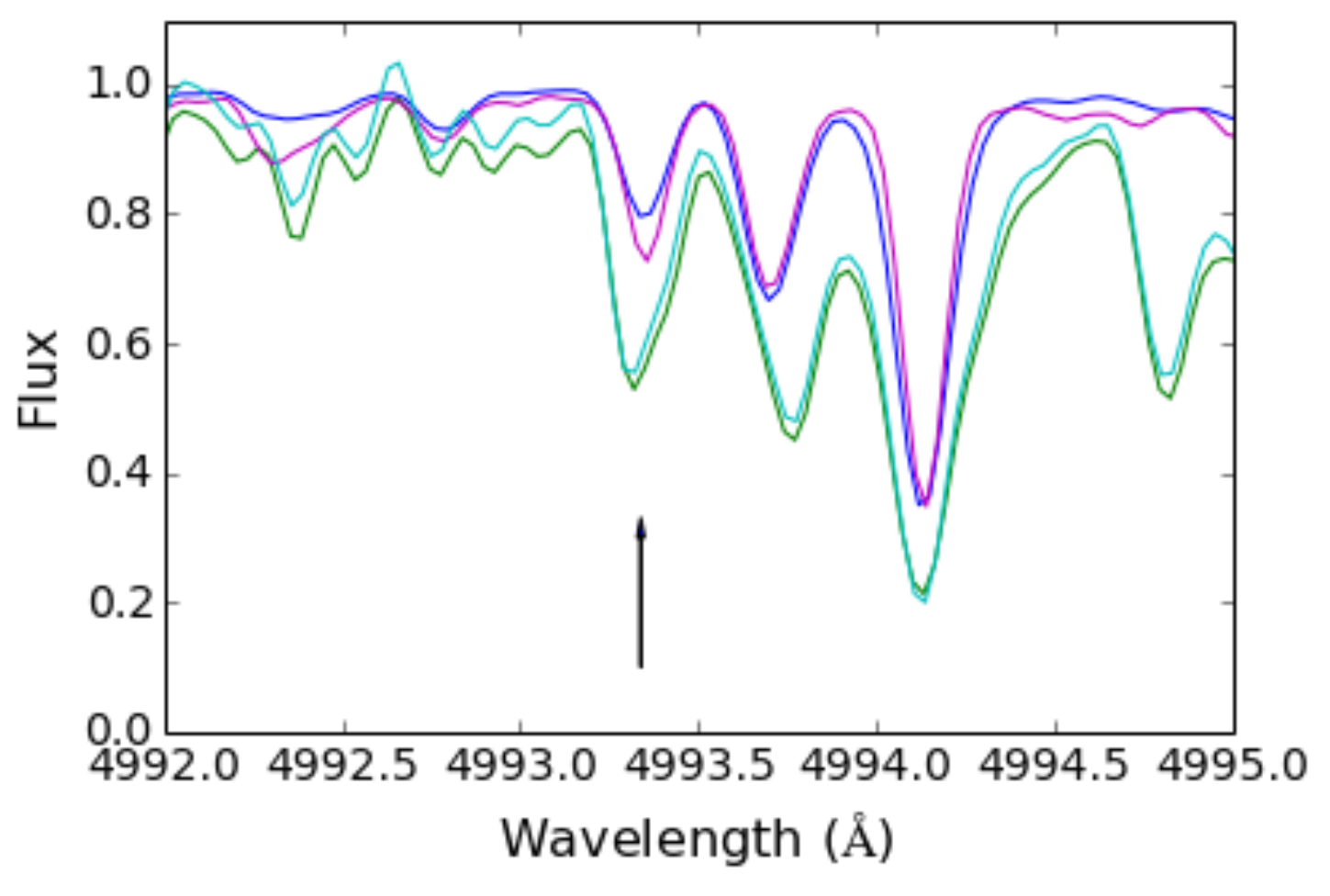}
    \caption{The spectra of binary stars HD131156a (dark blue), HD131156b (green), HD138004a (magenta), and HD138004b (cyan), with the arrow pointing at the region of interest: the ``bad" Fe {\scriptsize{II}} line at $\lambda$ = 4993.34 \AA.}
    \label{BinBadFeII}
  \end{minipage}
\end{figure}

Figure \ref{BinGoodFeII} depicts the ``good" Fe {\scriptsize{II}} line at $\lambda$ = 4491.4 \AA, where the spectra for HD131156a is in dark blue, HD131156b is in green, HD138004a is in magenta, and HD138004b is in cyan.

This figure is indicative of what we would expect from such a line: all four are centered, with a clean Gaussian profile, and are not blended with surrounding lines. The primary stars have deeper, stronger EWs, and the secondary stars have less excitation, leading to weaker Fe {\scriptsize{II}} lines. \par

Now let us see what happens when we inspect one of the ``bad" Fe {\scriptsize{II}} lines, at $\lambda$ = 4993.34 \AA, as depicted in Figure \ref{BinBadFeII}:

\begin{figure}[!t]
  \centering
    \begin{minipage}[b]{0.45\textwidth}
    \includegraphics[width=\textwidth]{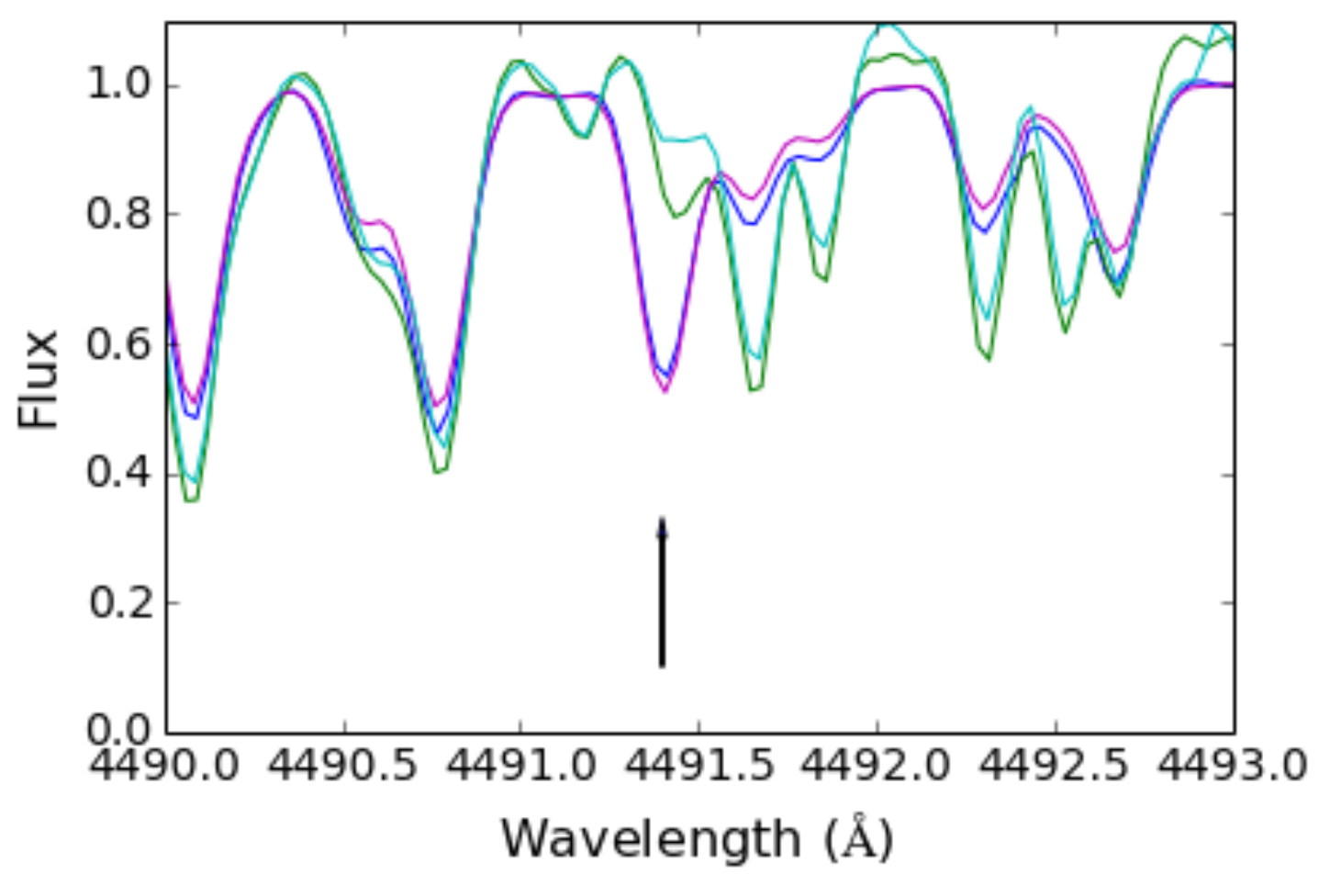}
    \caption{The spectra of Hyades stars HIP20146 (dark blue), HIP22177 (green), HIP19786 (magenta), and HIP19316 (cyan), with the arrow pointing at the region of interest: the ``good" Fe {\scriptsize{II}} line at $\lambda$ = 4491.4 \AA.}
    \label{HBinGood}
  \end{minipage}%
  \hfill
  \begin{minipage}[b]{0.45\textwidth}
    \includegraphics[width=\textwidth]{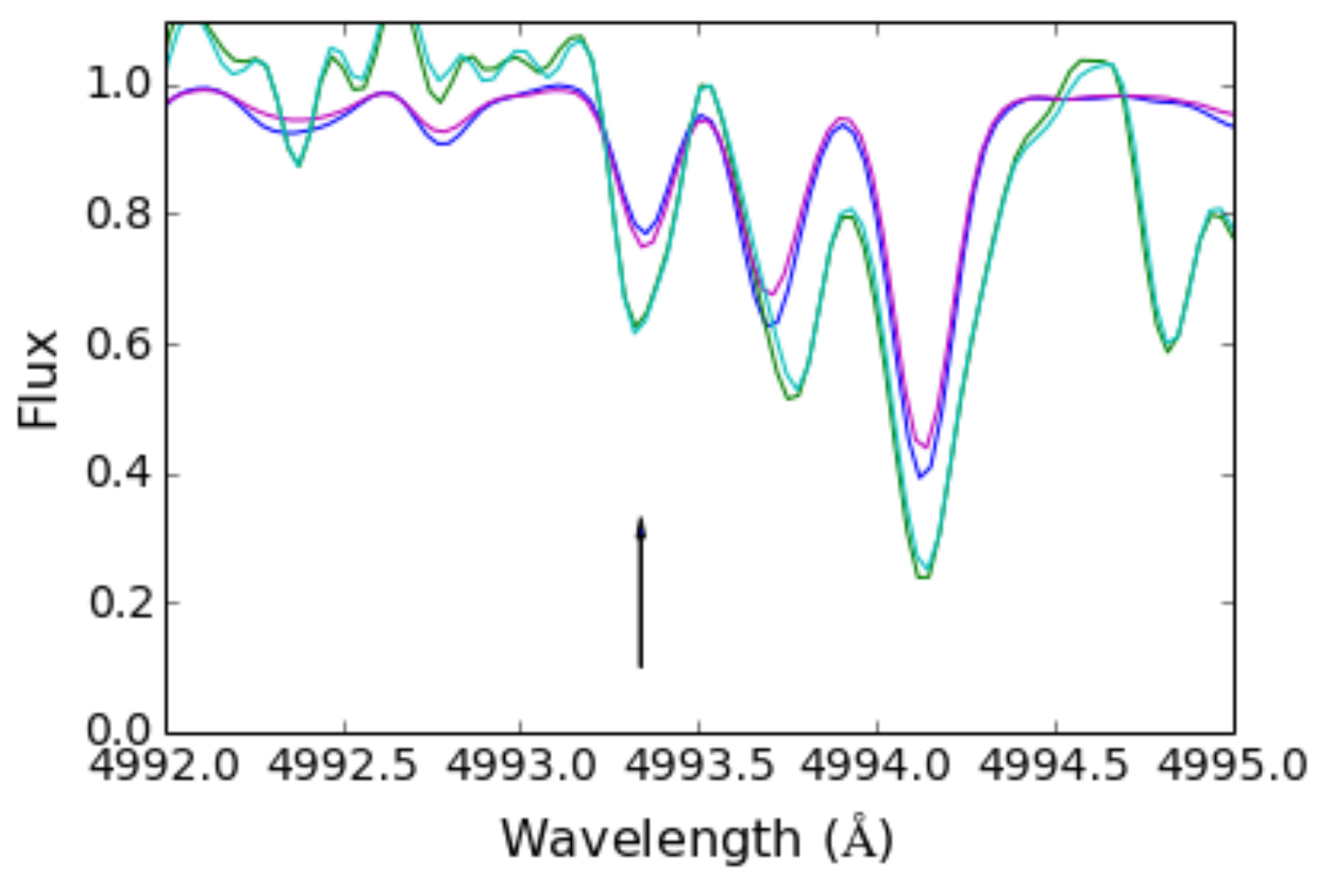}
    \caption{The spectra of Hyades stars HIP20146 (dark blue), HIP22177 (green), HIP19786 (magenta), and HIP19316 (cyan), with the arrow pointing at the region of interest: the ``bad" Fe {\scriptsize{II}} line at $\lambda$ = 4993.34 \AA.}
    \label{HBinBad}
  \end{minipage}
\end{figure}

Here, we immediately notice the difference. For the cooler K-dwarfs, the profiles are not centered on the precise wavelength, and the profile does not represent a pure Gaussian. Even when employing IRAF's \verb|deblend| command, we find this issue very difficult to resolve. Further, it is now the cooler stars that have the stronger, deeper profiles, indicative of line-blending affecting the line profiles more significantly in K-dwarfs.  

These results are analogous to what we find in the Hyades, found in Figures \ref{HBinGood} and \ref{HBinBad}, respectively. The ``good" Fe {\scriptsize{II}} line at $\lambda$ = 4491.4 {\AA} shows a nicely centered line with a strong Gaussian profile. Although HIP19316 and HIP22177 do not exhibit such clean profiles as their cool-temperature counterparts in the binaries, this is not a cause for concern; the Hyades is a young cluster, likely on average far younger than our binary field stars. Due to their young age, they exhibit rotation and magnetic activity which widens the profile more so than in the binaries. In regard to the ``bad" Fe {\scriptsize{II}} line at $\lambda$ = 4993.34 \AA, we see the same uneven Gaussian profile, with their troughs misaligned. Because this and five other Fe {\scriptsize{II}} lines had similar problems, and consistently gave suspiciously high abundances, we removed them from our linelist. \par

\begin{figure}[H]
\centering
\includegraphics[width=1\linewidth]{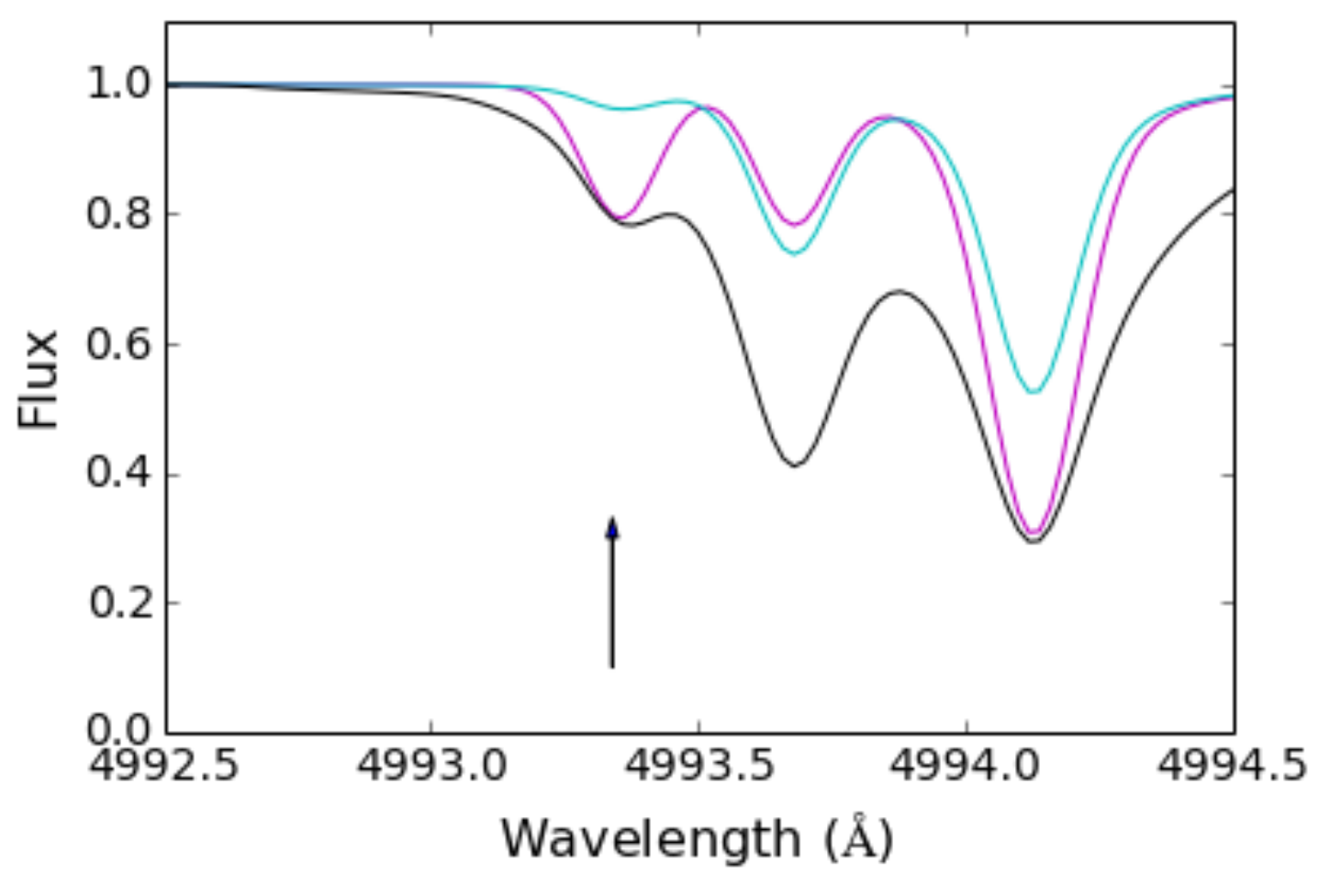}
\vspace{-0.3cm}
\caption{The synthetic spectrum calculation of binary star HD138004a (magenta), and HD138004b (cyan), with the arrow pointing at the region of interest: the ``bad" Fe {\scriptsize{II}} line at $\lambda$ = 4993.34 \AA. Another synthetic spectrum for HD138004b where [Fe/H] is increased by 1.2 dex in order to match the synthetic spectrum for HD138004a is shown in black.}
\label{abd}
\end{figure}

To further justify removing these lines, we performed synthetic spectral line calculations to estimate the degree of blending. Synthetic model atmospheres were created using the MARCS grid (for linear interpolation grid consistency) for the star, as well as its $T_\mathrm{eff}$, log $g$, [Fe/H] and $\xi$. Using MOOG \citep{Sneden73}, we added atomic spectra data from the NIST Database\footnote{https://www.nist.gov/pml/atomic-spectra-database} for the lines which had values for $\lambda$ around the blended line of interest, species classification, excitation potential $\chi$ $\leq$ 10 eV (as anything with $\textgreater$ 10 eV is unlikely to appear in spectra of cool stars), and transition strength in log $gf$. In the synthesis, we adopted a Gaussian smoothing of 0.15 \AA\ of FWHM (full width at half maximum) to simulate the total line broadening. Here in Figure \ref{abd} we recreate part of Figure \ref{BinBadFeII} for HD138004a and HD138004b only with the synthetic spectrum, using the same labels for clarity.

According to our synthesis, the desired blended line at $\lambda$ = 4993.34 \AA\ in Figure \ref{BinBadFeII} should be much weaker in HD138004b than in HD138004a, but in fact we find the opposite. An analogous synthesis was performed for HD131156b/HD131156a (not shown), and an identical result was found. Although it is not our purpose to match the observed spectra, but rather make an order-of-magnitude estimate of these effects, it is still reasonable to expect the Fe {\scriptsize{II}} line to be weaker in the K-dwarfs HD138004b and HD131156b. This is especially true when we consider how the predicted flux of the line core matched well for HD138004a and HD131156a in both our observed and synthetic spectrum: $\thicksim$0.8 at $\lambda$ = 4993.34 \AA, $\thicksim$0.7 at $\lambda$ = 4993.68 \AA, and $\thicksim$0.3 at $\lambda$ = 4994.129 \AA. However, we predicted the flux for both K-dwarfs to be $\thicksim$0.97 at $\lambda$ = 4993.34 \AA\ but observed a flux of $\thicksim$0.55. Meanwhile we found that the Fe {\scriptsize{I}} line $\lambda$ = 4993.68 \AA\ in the K-dwarfs have a more deep Gaussian profile in both our observed and theoretical spectra, alluding that our theoretical calculations should preserve whether the solar type or K-dwarfs have the deeper trough. Thus, the Fe {\scriptsize{II}} line at $\lambda$ = 4993.34 \AA\ (and subsequently the Fe {\scriptsize{I}} line at $\lambda$ = 4994.129 \AA, but that is beyond the scope of this work) is likely predominantly affected by line-blends from atomic and molecular lines. \par

We further tried to quantify the degree of blending by artificially increasing the iron abundance in our secondary star to match the theoretical spectrum of its primary partner. For stars HD138004b ($T_\mathrm{eff}$ = 4327 K) and HD131156b ($T_\mathrm{eff}$ = 4359 K), we found that we needed to artificially increase [Fe/H] by 1.2 dex (that is, 1.2 dex higher than the abundance of the model atmosphere) to match the theoretical spectra of HD138004a and HD131156a at $\lambda$ = 4993.34 \AA. Likewise, for another blended line identified by this present work at $\lambda$ = 4508.29 \AA, we find that we need an increased abundance of 1.3 dex. To see how this varied with temperature, we retried this with some slightly warmer secondaries: HD69056b ($T_\mathrm{eff}$ = 4663 K) and HD118576b ($T_\mathrm{eff}$ = 5041 K). At $\lambda$ = 4993.34 \AA\ as well as $\lambda$ = 4508.29 \AA, we only needed to increase [Fe/H] by 0.9 to best match HD69056a and by 0.5 to best match HD118576a. It must be emphasized that we are not deriving an abundance from the synthesis, but only showing the behavior of Fe {\scriptsize{II}} lines and their dependency on $T_\mathrm{eff}$. \par

From the synthetic spectrum, the behavior is clear: the intensity of the iron lines decreases with decreasing temperature, and so the K-dwarfs need increasing artificial abundance to make up for this deficit. This trend further indicates that stars are more heavily affected by blends with decreasing temperature. \par

We acknowledge that these calculations were performed without modeling molecular lines. Including them in the calculation would lower the overall level of the continuum and potentially add lines on top of the Fe {\scriptsize{II}} feature, rendering our estimates possibly too large, but this is somewhat irrelevant for our purpose: to prove with synthetic spectra that the Fe {\scriptsize{II}} lines alone decrease in strength as $T_\mathrm{eff}$ decreases. In this exercise we only care about how Fe {\scriptsize{II}} lines, when analyzed alone, are not being correctly modeled.

Thus, our final linelist is presented in Table \ref{linelist}, below, where a $\ddag$ represents a deleted line due to line blends. It must be noted, however, that the Fe {\scriptsize{II}} linelist was constructed using the solar spectrum as reference, meaning that they are likely not as useful for K-dwarfs but are acceptable for Sun-like stars.

\FloatBarrier
\raggedbottom
\setlength{\LTpre}{0pt}
\begin{center}
\begin{longtable}{lllrrr}
\nopagebreak
\caption{\label{linelist} Atomic Line List}\\
\hline\hline
Wavelength (\AA) & $\chi$ (eV) & log $gf$ \\ 
\hline
Fe {\scriptsize{I}} \\
\hline
\endfirsthead
\caption{continued.}\\
\hline\hline
Wavelength (\AA) & $\chi$ (eV) & log $gf$ \\ 
\hline
\endhead
\hline
\endfoot
4389.25 & 0.05 & -4.58  \\ 
4445.47 & 0.09 & -5.44  \\ 
4602.00 & 1.61 & -3.15 \\
4690.14 & 3.69 & -1.61 \\
4788.76 & 3.24 & -1.73 \\
4799.41 & 3.64 & -2.13 \\
4808.15 & 3.25 & -2.69 \\
4950.10 & 3.42 & -1.56 \\
4994.13 & 0.92 & -3.08 \\
5141.74 & 2.42 & -2.23 \\
5198.71 & 2.22 & -2.14 \\
5225.53 & 0.11 & -4.79 \\
5242.49 & 3.63 & -0.99 \\
5247.05 & 0.09 & -4.96 \\
5250.21 & 0.12 & -4.94 \\
5295.31 & 4.42 & -1.59 \\
5322.04 & 2.28 & -2.89 \\
5373.71 & 4.47 & -0.74 \\
5379.57 & 3.69 & -1.51 \\
5386.33 & 4.15 & -1.67 \\
5441.34 & 4.31 & -1.63 \\
5466.40 & 4.37 & -0.57 \\
5466.99 & 3.57 & -2.23 \\
5491.83 & 4.19 & -2.19 \\
5554.89 & 4.55 & -0.36 \\
5560.21 & 4.43 & -1.09 \\
5618.63 & 4.21 & -1.27 \\
5638.26 & 4.22 & -0.77 \\
5651.47 & 4.47 & -1.75 \\
5679.02 & 4.65 & -0.75 \\
5701.54 & 2.56 & -2.16 \\
5705.46 & 4.30 & -1.36 \\
5731.76 & 4.26 & -1.20 \\
5775.08 & 4.22 & -1.30 \\
5778.45 & 2.59 & -3.44 \\
5784.66 & 3.40 & -2.53 \\
5793.91 & 4.22 & -1.62 \\
5806.73 & 4.61 & -0.95 \\
5852.22 & 4.55 & -1.23 \\
5855.08 & 4.61 & -1.48 \\
5930.18 & 4.65 & -0.17 \\
5934.65 & 3.93 & -1.07 \\
5956.69 & 0.86 & -4.55 \\
5987.07 & 4.80 & -0.21 \\
6003.01 & 3.88 & -1.06 \\
6005.54 & 2.59 & -3.43 \\
6027.05 & 4.08 & -1.09 \\
6056.00 & 4.73 & -0.40 \\
6065.48 & 2.61 & -1.53 \\
6079.01 & 4.65 & -1.02 \\
6082.71 & 2.22 & -3.57 \\
6093.64 & 4.61 & -1.30 \\
6096.67 & 3.98 & -1.81 \\
6151.62 & 2.18 & -3.28 \\
6165.36 & 4.14 & -1.46 \\
6173.34 & 2.22 & -2.88 \\
6187.99 & 3.94 & -1.62 \\
6200.31 & 2.61 & -2.42 \\
6213.43 & 2.22 & -2.52 \\
6219.28 & 2.20 & -2.43 \\
6226.74 & 3.88 & -2.10 \\
6232.64 & 3.65 & -1.22 \\
6240.65 & 2.22 & -3.29 \\
6265.13 & 2.18 & -2.55 \\
6271.28 & 3.33 & -2.70 \\
6322.69 & 2.59 & -2.43 \\
6380.74 & 4.19 & -1.32 \\
6392.54 & 2.28 & -4.03 \\
6430.85 & 2.18 & -2.01 \\
6498.94 & 0.96 & -4.70 \\
6593.87 & 2.43 & -2.39 \\
6597.56 & 4.80 & -0.97 \\
6625.02 & 1.01 & -5.34 \\
6703.57 & 2.76 & -3.02 \\
6705.10 & 4.61 & -0.98 \\
6710.32 & 1.49 & -4.88 \\
6713.75 & 4.80 & -1.40 \\
6725.36 & 4.10 & -2.19 \\
6726.67 & 4.61 & -1.03 \\
6733.15 & 4.64 & -1.47 \\
6739.52 & 1.56 & -4.79 \\
6750.15 & 2.42 & -2.62 \\
6806.85 & 2.73 & -3.11 \\
6810.26 & 4.61 & -0.99 \\
6837.01 & 4.59 & -1.69 \\
6839.83 & 2.56 & -3.35 \\
6843.66 & 4.55 & -0.83 \\
6858.15 & 4.61 & -0.94 \\
7583.79 & 3.02 & -1.88 \\
\hline
Fe {\scriptsize{II}} \\
\hline
4491.40 & 2.86 & -2.66 \\
4508.29\ddag & 2.86 & -2.52 \\
4576.33 & 2.84 & -2.95 \\
4620.51 & 2.83 & -3.21 \\
4993.34\ddag & 2.81 & -3.73 \\
5197.58\ddag & 3.23 & -2.22 \\
5234.62 & 3.22 & -2.18 \\
5264.80 & 3.23 & -3.13 \\
5325.55\ddag & 3.22 & -3.25 \\
5414.07 & 3.22 & -3.58 \\
5425.26\ddag & 3.20 & -3.22 \\
6084.09 & 3.20 & -3.83 \\
6149.24 & 3.89 & -2.75 \\
6247.55 & 3.89 & -2.38 \\
6369.46 & 2.89 & -4.11 \\
6432.68 & 2.89 & -3.57 \\
6456.38\ddag & 3.90 & -2.05 \\
7515.83 & 3.90 & -3.39 \\
\hline
\end{longtable}
\end{center}
\setlength{\LTpost}{-10pt}
\FloatBarrier
\vspace{-0.7cm}

As there were a total of only 18 Fe {\scriptsize{II}} lines originally, it is natural to reason that 6 bad lines would greatly skew Fe {\scriptsize{II}} measurements. Earlier, we mentioned the analysis of measured Fe {\scriptsize{I}} and Fe {\scriptsize{II}} lines in the Hyades cluster from \cite{Yong04}. In this, they did not account for line-blending effects. In fact, they used four of the six lines we determined to be blended, specifically $\lambda$ = \{4508.29 \AA, 5197.58 \AA, 5325.55 \AA, 5425.26 \AA\}. \par

In their analysis, \cite{Yong04} plotted their measured Fe {\scriptsize{I}} and Fe {\scriptsize{II}} abundances as a function of $T_\mathrm{eff}$. They, too, discovered a statistically significant offset in Fe {\scriptsize{I}} vs Fe {\scriptsize{II}} predictions. At $T_\mathrm{eff}$ $\textgreater$ 5000 K, they found $\Delta$[Fe/H] = 0.02 dex, but this increased with decreasing $T_\mathrm{eff}$. In the coolest temperature stars at $T_\mathrm{eff}$ = 4000 K they found a disagreement of 1.0 dex. Their inferred iron abundances from Fe {\scriptsize{I}} alone were approximately constant in stars with $T_\mathrm{eff}$ $\textgreater$ 4300 K, but increases to about 0.2 dex in stars of $T_\mathrm{eff}$ = 4000 K. \par

\begin{figure}[H]
\centering
\includegraphics[width=1\linewidth]{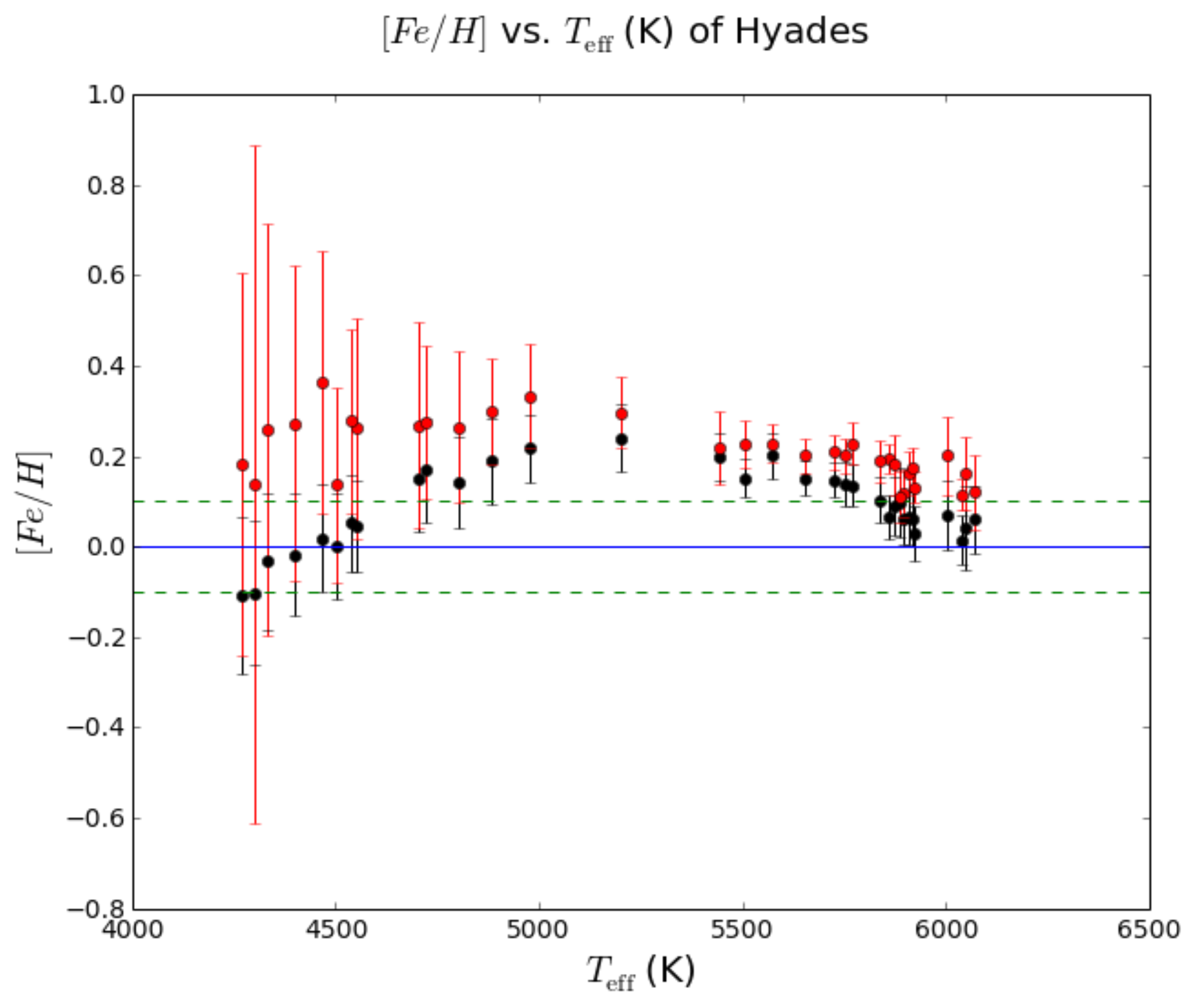}
\vspace{-0.2cm}
\caption{Mean iron abundance of Hyades stars derived from Fe {\scriptsize{I}} (black circles) and Fe {\scriptsize{II}} (red circles) lines separately. The green dashed lines represent $\pm$0.1 dex.}
\label{Yongfigure}
\end{figure}

In comparison, we replicated this plot with our own Hyades data, this time neglecting the blended lines, as seen in Figure \ref{Yongfigure}. Unlike their plot, iron abundances as inferred from both Fe {\scriptsize{I}} (black circles) and Fe {\scriptsize{II}} (red circles) lines follow a slight parabolic shape, with an approximate maximum of [Fe/H] = 0.3 for Fe {\scriptsize{II}} and [Fe/H] = 0.2 for Fe {\scriptsize{I}}. In the middle at $T_\mathrm{eff}$ = 5200 K, where this approximate maximum occurs, both lines decrease by about 0.2-0.3 dex with both warmer and cooler temperatures, save for some Fe {\scriptsize{II}} lines below 5200 K which remain fairly constant, dropping on average by 0.1 dex. This parabolic shape was not seen in the plot from \cite{Yong04}. In fact, we observe in our sample that [Fe/H] as inferred from Fe {\scriptsize{I}} lines decrease with temperatures below 5200 K, which is the opposite of their findings. A possible explanation comes from the method with which $T_\mathrm{eff}$ was derived; \cite{Yong04} used the $T_\mathrm{eff}$:[Fe/H]:color relations based on the infrared flux method from \cite{Alonso96}. For stars with Str\"{o}mgren $b-y$ index and $B-V$ photometry, they adopted the mean $T_\mathrm{eff}$. We derive $T_\mathrm{eff}$ from direct measurements using LDR, and when we compared the differences in $T_\mathrm{eff}$ between our samples ($T_\mathrm{eff, Yong}$ - $T_\mathrm{eff, Aleo}$), we found that they underestimated the temperature by $\thicksim$ 250 K in the coolest stars and overestimated the temperature by $\thicksim$ 200 K in the warmest stars. Our calculations match at $\thicksim$ 5500 K, which is approximately the location of the maximum [Fe/H] in our plot.

After removing the six identified ``bad" lines in our sample, one would expect to see the abundance anomaly corrected, despite some star to star variations.  \par

\begin{figure*}[!t]
   \centering
    \begin{minipage}[b]{\textwidth}
    \centering
    \includegraphics[width=14cm, height=11cm, trim= 0 0 0 0]{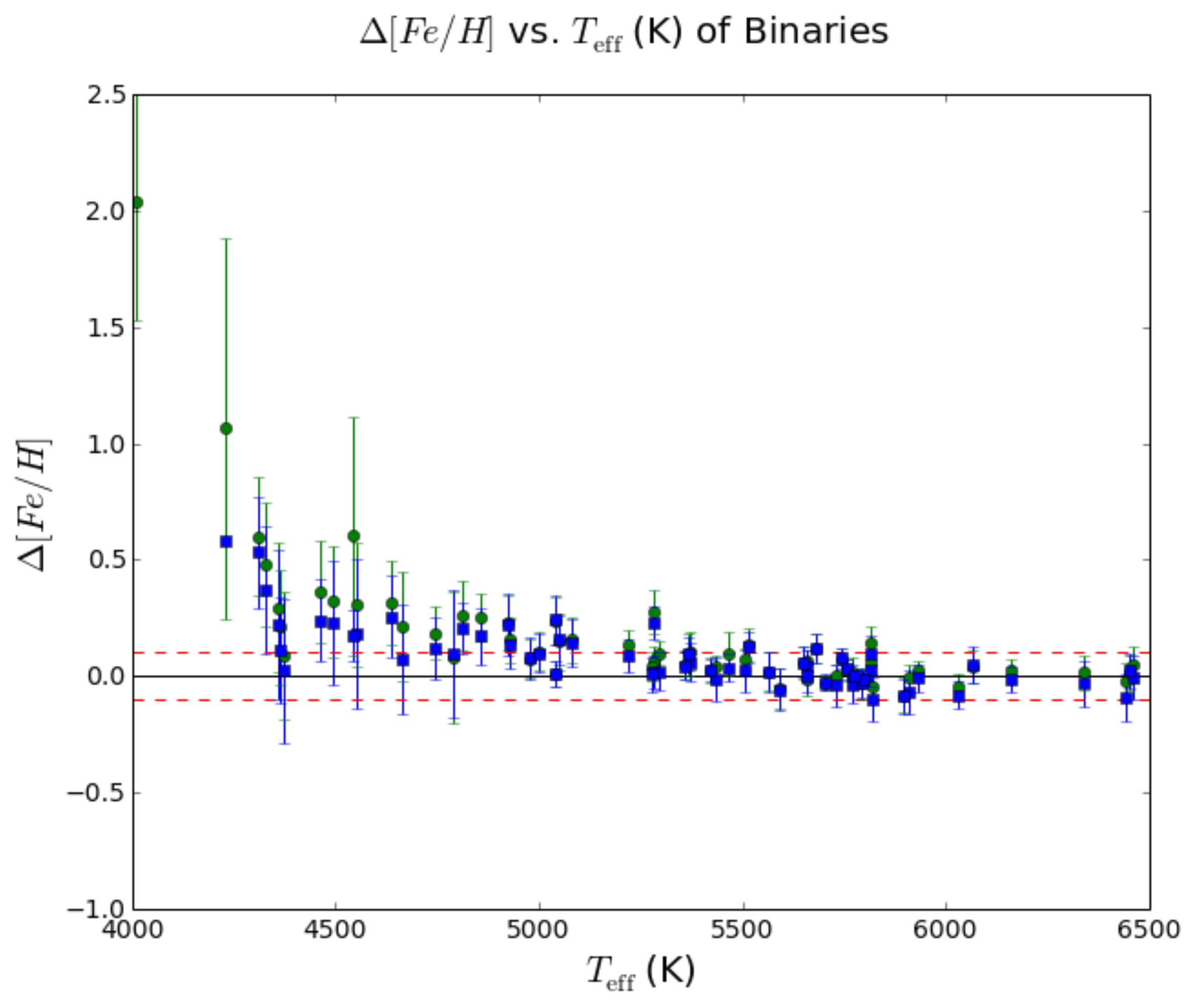}
    \caption{The difference in iron abundance (Fe {\scriptsize{II}} - Fe {\scriptsize{I}}) of our binary star sample. The green circles represent our initial measurements before the 6 Fe {\scriptsize{II}} bad lines are removed from our calculations. The blue squares are the results after the removal of these lines. The red dashed lines represent $\pm$0.1 dex.}
    \label{beforeandafter}
  \end{minipage}%
  \hfill
 \centering
  \begin{minipage}[b]{\textwidth}
  \centering
    \includegraphics[width=14cm, height=11cm, trim= 0 0 0 0]{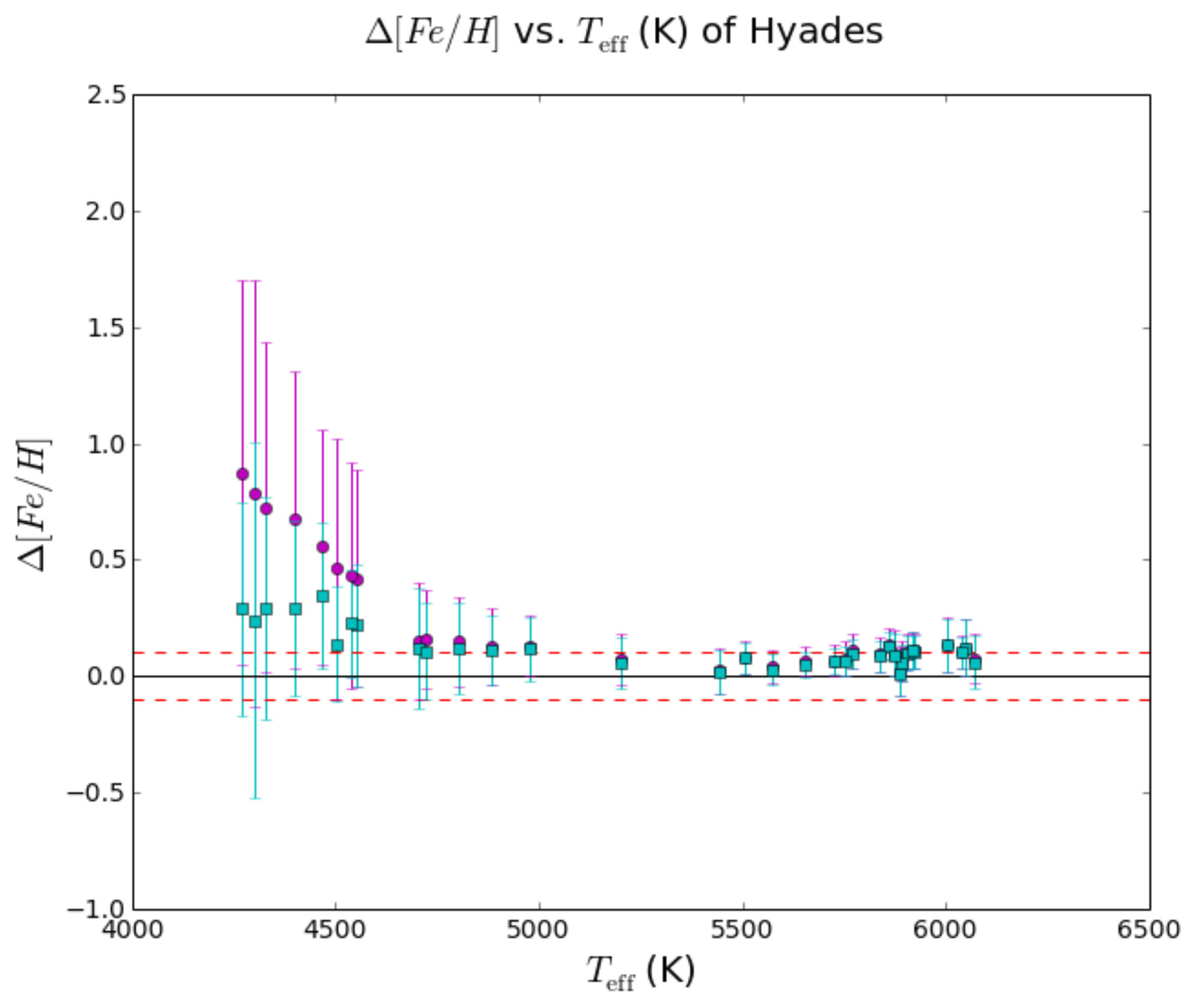}
     \caption{The difference in iron abundance (Fe {\scriptsize{II}} - Fe {\scriptsize{I}}) of our Hyades star sample. The magenta circles represent our initial measurements before the 6 Fe {\scriptsize{II}} bad lines are removed from our calculations. The cyan squares are the results after the removal of these lines. The red dashed lines represent $\pm$0.1 dex.}
    \label{hyadesbeforeandafter}
  \end{minipage}
\end{figure*}

However, this is not entirely the case. When plotting the \textit{difference} in iron abundance (Fe {\scriptsize{II}} - Fe {\scriptsize{I}}) of our binary and Hyades star samples, the trend is severely depreciated, but not completely corrected. Figure \ref{beforeandafter} depicts our abundance measurements before (the green circles) and after (blue squares) the removal of the ``bad" lines for our binary sample. We applied this process to our Hyades sample as well, with magenta circles representing our abundances before the removal of the ``bad" lines and cyan squares after. This is shown in Figure \ref{hyadesbeforeandafter}.

In Figure \ref{beforeandafter}, it is evident that the removal of our blended lines impacts our trend, as high as a full order of magnitude in some cases. Now most of the stars, accounting for error, have $\Delta$[Fe/H] within 0 or 0.1 dex. Despite this correction, a noticeable trend is still present below a $T_\mathrm{eff}$ of about 4700 K in both samples. Considering the contribution of this temperature range to our anomaly, we shall reference the stars cooler than $T_\mathrm{eff} = 4700$ K as ``trend" stars. Even after the removal the ``bad" lines, there is still a lingering factor or two which is forcing overionization in these ``trend" stars. 

It is also apparent that the removal of our blended lines did not completely rid our Hyades sample (Figure \ref{hyadesbeforeandafter}) of the abundance anomaly, most notably in our ``trend" stars. Although it improved $\Delta$[Fe/H] by about 0.6 dex at maximum, there is still an offset of about 0.3 dex that cannot be explained by line-blending. Let us investigate other possible factors.

%%%%%%%%%%%%%%%%%%%%%
%%%%%%%%%%%%%%%%%%%%%

\subsection{Possible Departures from LTE}

For this section, let us posit that NLTE effects are to blame. If so, a prime candidate causing the overabundance is the flux of ultraviolet photons penetrating line-forming regions \citep{Yong04}. The resulting calculations of this nature have been investigated in approximately solar-like stars ([Fe/H] $\thicksim$ 0, $T_\mathrm{eff}$ $\thicksim$ 5777 K and greater), and the consensus is that these effects are small, taking place mostly in granular atmospheric regions \citep{Gehren01, Shchukina01}. In \cite{Shchukina01}, they find that the NLTE fit to observed EW is about 0.074 dex larger than for LTE, and that the largest differences are found for the weakest low-excitation Fe {\scriptsize{I}} lines of excitation potential $\textless$ 2 eV. This is non-neglible, but it in no way would account for our total observed difference. As a whole, the ionization due to near-UV radiation leads to a small over-excitation in the Fe {\scriptsize{II}} levels relative to LTE and an underpopulation of the Fe {\scriptsize{I}} levels. This is mainly due to the mean energy of the flux of these near-UV photons ($J_{\nu}$) exceeding the flux assumed in LTE from the Planck function ($B_{\nu}$). In effect, this could spike abundance of Fe {\scriptsize{II}} lines while only slightly increasing abundance in Fe {\scriptsize{I}}, as would be the case if this was a large effect at these cooler temperatures. However, granulation is not as important in K-dwarfs as they are in solar temperature stars, leading us to believe that this NLTE mechanism is not a significant fraction of the difference we observe.  \par

Recently, more NLTE calculations have been done in our desired parameter range. \cite{Sitnova15} performed a systematic NLTE study of Fe {\scriptsize{I}} and Fe {\scriptsize{II}} in -2.6 $\leq$ [Fe/H] $\leq$ 0.2 F and G dwarfs in the solar neighborhood. They found abundances from the two ionization stages to be consistent within 0.06 dex for every star, when applying a scaling factor $S$\textsubscript{H} = 0.5 to Drawinian rates of inelastic Fe+H collisions. In fact, the difference in average abundance between NLTE and LTE was less than 0.06 dex for stars with either [Fe/H] $\geq$ -0.75, or $T_\mathrm{eff}$ $\leq$ 5750 K, or log $g$ $\geq$ 4.20. Their finding of 0.06 dex is small, and would in no way lead to the difference of up to 0.3 dex that we observe in the Hyades and the 0.6 dex we observe in the binaries. \par

In a similar parameter range of 4000 K $\leq$ $T_\mathrm{eff}$ $\leq$ 5000 K and a metal abundance of -4 $\leq$ [Fe/H] $\leq$ 0, \cite{Mashonkina16} found the departures from LTE are small in the entire range of stellar parameters under consideration. Specifically, they discovered $\Delta$NLTE does not exceed 0.01 dex in absolute value for all 20 Fe {\scriptsize{II}} lines in all their models. Likewise, adopting 1D model atmospheres in another sample composed of the Sun and five reference stars with reliable stellar parameters of 4600 K $\leq$ $T_\mathrm{eff}$ $\leq$ 6500 K, 1.6 $\leq$ log $g$ $\leq$ 4.53, -2.5 $\leq$ [Fe/H] $\leq$ 0.1, \cite{Mashonkina11} found that the departures from LTE are again negligible for the Fe {\scriptsize{II}} lines over the whole stellar parameter range considered. \par

In speaking privately with Mashonkina about the stars in our sample, she writes that the departures from LTE increase towards higher $T_\mathrm{eff}$ and lower log $g$. In this case, such as in the Sun, Procyon, and beta Vir, the NLTE abundance corrections reach no more than -0.01 dex, adding that NLTE effects for Fe {\scriptsize{II}} are expected to be even smaller for sunspots because of their lower $T_\mathrm{eff}$, rendering NLTE effects for our K-dwarfs of interest to also be negligible. \par

\cite{Lind12} found similar results, but for Fe {\scriptsize{I}}. Specifically, NLTE calculations show Fe {\scriptsize{I}} lines are increasingly underestimated in hotter, lower surface gravity and more metal-poor stars, whereas LTE is usually a ``realistic approximation" for Fe {\scriptsize{II}} lines. They also found that the 3D model-derived average metallicities are not dramatically different from 1D models in NLTE ($\leq$ 0.04 dex), which implies that NLTE effects on our 1D-LTE model atmospheres should be mostly consistent, despite the dimension of NLTE models. \par

Due to these recent findings, there is considerable doubt that NLTE effects are responsible for the resulting trend we observe.

To convolute the issue, the Hyades dwarfs are chromospherically active \citep{Duncan84, Reid95}, which would imply for greater NLTE effects \citep{Yong04}. On average, our binary sample should be no more active than these Hyades dwarfs, yet the trend is not as prevalent in our Hyades sample as in our binary sample. \par

%%%%%%%%%%%%%%%%%%%%%
%%%%%%%%%%%%%%%%%%%%%

\subsection{Age \& Activity}

\begin{figure}[H]
\centering
\includegraphics[width=1\linewidth]{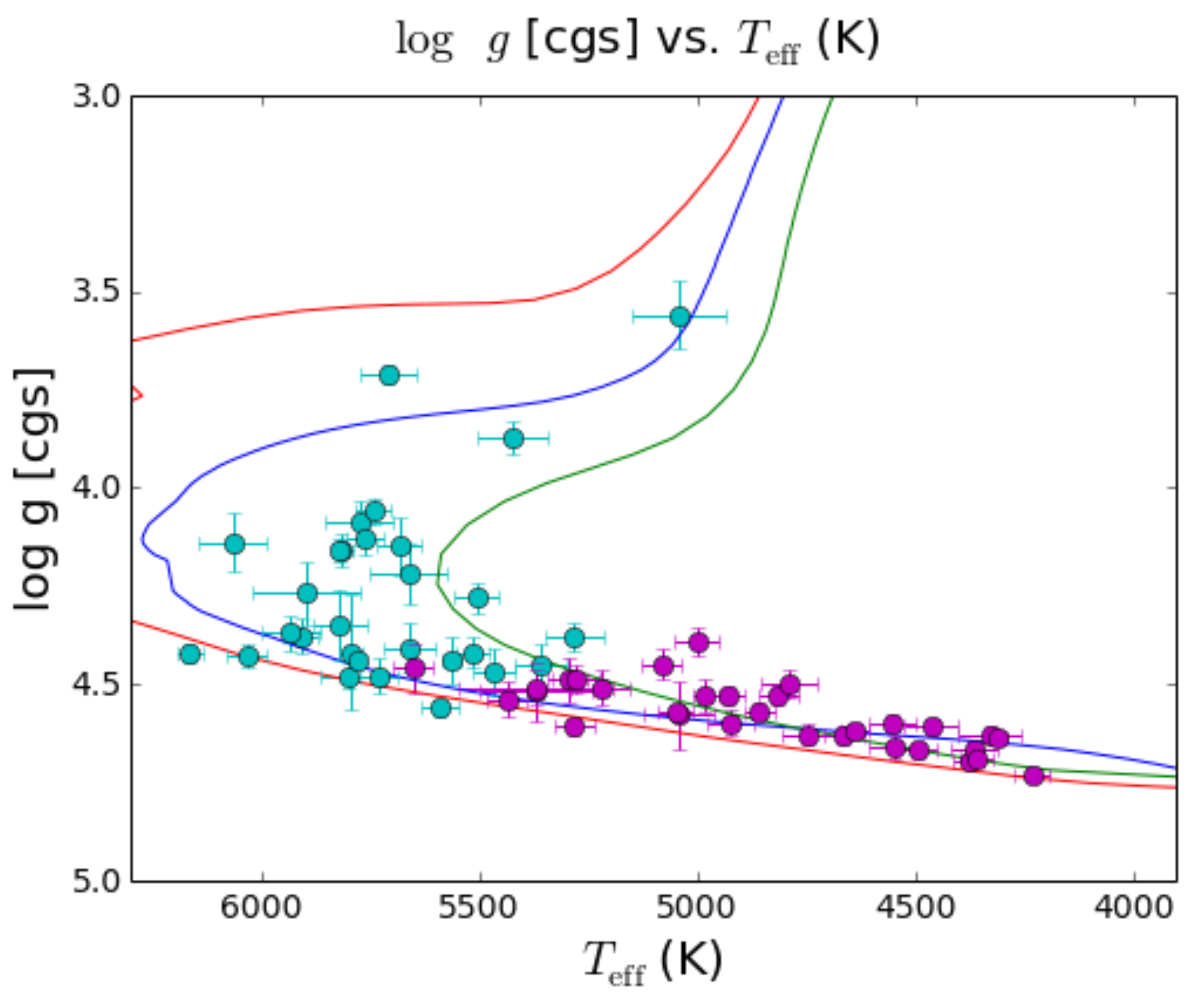}
\vspace{-0.3cm}
\caption{Our primary stars (cyan) are plotted with our secondary stars (magenta) on three isochrones: the green isochrone is of age 12.0 Gyr, the blue isochrone is of solar age, and the red isochrone is 2.0 Gyr. Errors in the stars' $T_\mathrm{eff}$ and log $g$ are represented by the cross-hairs.}
\label{HR}
\end{figure}

Since we can reasonably rule out NLTE effects as the cause for our abundance problem, we now investigate other culprits. It is possible that chromospheric activity, which is strongly related to stellar age, could prove to be a factor. Combining our isochrone data from the Dartmouth Stellar Evolution Database with our most updated star parameters using \verb|astropy.io|, we created a diagram to illustrate log $g$ vs. $T_\mathrm{eff}$ of our binary stars, as seen in Figure \ref{HR}.

Here we overlaid three isochrones onto this diagram, where the green isochrone is of age 12.0 Gyr, the blue isochrone is of solar age, and the red isochrone is 2.0 Gyr. With this, we can obtain a general sense of the ages of our starlist. This is more easily deduced using primary stars, but these are not the focus of our experiment. \par

Since we are focusing on the secondary stars, we want to see if there is any correlation with these stars and age, as it is reasonable to conjecture that the youngest, most active stars exhibit the abundance discrepancy. Unfortunately, the isochrones all converge at lower $T_\mathrm{eff}$, and this, compounded with the error of both log $g$ and $T_\mathrm{eff}$, makes it difficult to discern the true ages of our secondary stars and identify a correlation, assuming there is one. Due to this unreliability, we investigated the Ca {\scriptsize{II}} H+K emission and Li absorption lines, as the younger, more active stars will have more pronounced Ca {\scriptsize{II}} H+K emission and Li absorption. We admit that this connection is weak and very complex, but let us assume this correlation holds true. This way, we can group which of our secondary stars are more active/young and less active/old and see if this may result in the observed trend.

%%%%%%%%%%%%%%%%%%%%%%%
%%%%%%%%%%%%%%%%%%%%%%%

\subsubsection{Ca {\scriptsize{II}} H+K Emission}

We investigated the Ca {\scriptsize{II}} H 3968.5 {\AA} and the Ca {\scriptsize{II}} K 3933.6 {\AA} line. For each secondary (K-dwarf), we looked for the peak of the spectra relative to the normalized continuum of the spectra.  \par

In some cases the peak was clearly emission, in others it was absorption, and others it was a combination in which parts of the line were emission, but the peak would dip slightly. All of these cases were documented for all stars. We did not take explicit EW measurements, however these were not ultimately needed as we achieved a general sense as to which stars were older; it is known that a stronger emission in these Ca {\scriptsize{II}} H+K lines generally correlates with a younger stellar age. \par

Following this, we summed our measurements of the height of emission peaks for each star and subsequently averaged them. With this average, the stars were split into two groups: ones which have less emission than average (presumed to be the older stars in the group) and ones which which have more emission than average (presumed to be the younger stars in the group). The following figures \ref{CaHLess}, \ref{CaHGreater}, \ref{CaKLess}, \ref{CaKGreater} contain our results.

\begin{figure*}[!tbp]
  \centering
  \begin{minipage}[b]{0.45\textwidth}
    \includegraphics[width=\textwidth]{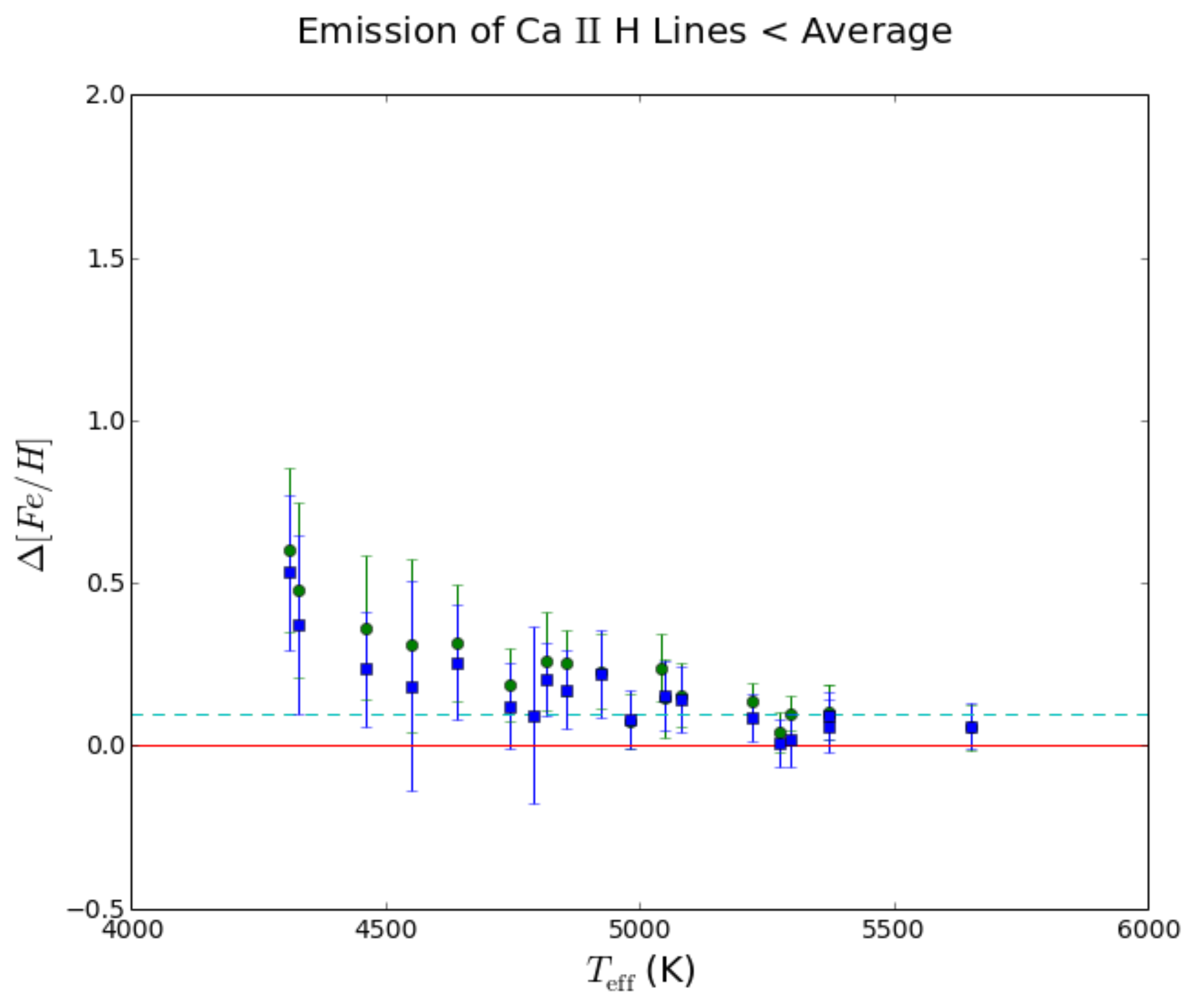}
    \caption{The difference in iron abundance (Fe {\scriptsize{II}} - Fe {\scriptsize{I}}) of our binary star secondaries where the emission as indicated by Ca {\scriptsize{II}} H 3968.5 {\AA} is \textit{less} than average when accounting for all stars in the sample. The green circles represent our initial measurements before the 6 Fe {\scriptsize{II}} bad lines are removed from our calculations. The blue squares are the results after the removal of these lines.}
    \label{CaHLess}
  \end{minipage}
  \hfill
  \begin{minipage}[b]{0.45\textwidth}
    \includegraphics[width=\textwidth]{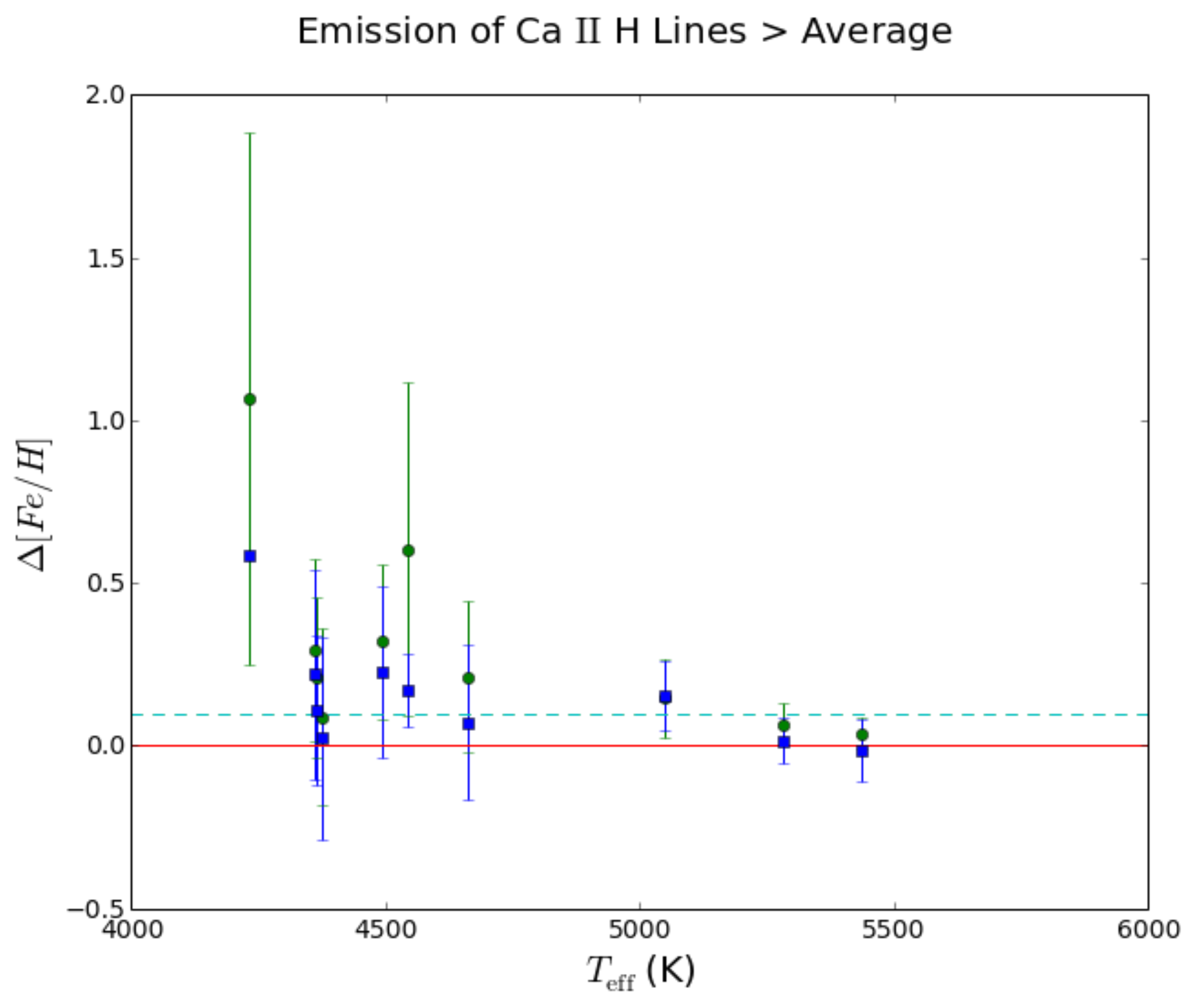}
    \caption{The difference in iron abundance (Fe {\scriptsize{II}} - Fe {\scriptsize{I}}) of our binary star secondaries where the emission as indicated by Ca {\scriptsize{II}} H 3968.5 {\AA} is \textit{greater} than average when accounting for all stars in the sample. The green circles represent our initial measurements before the 6 Fe {\scriptsize{II}} bad lines are removed from our calculations. The blue squares are the results after the removal of these lines.}
    \label{CaHGreater}
  \end{minipage}
\end{figure*}

\begin{figure*}[!tbp]
  \centering
  \begin{minipage}[b]{0.45\textwidth}
    \includegraphics[width=\textwidth]{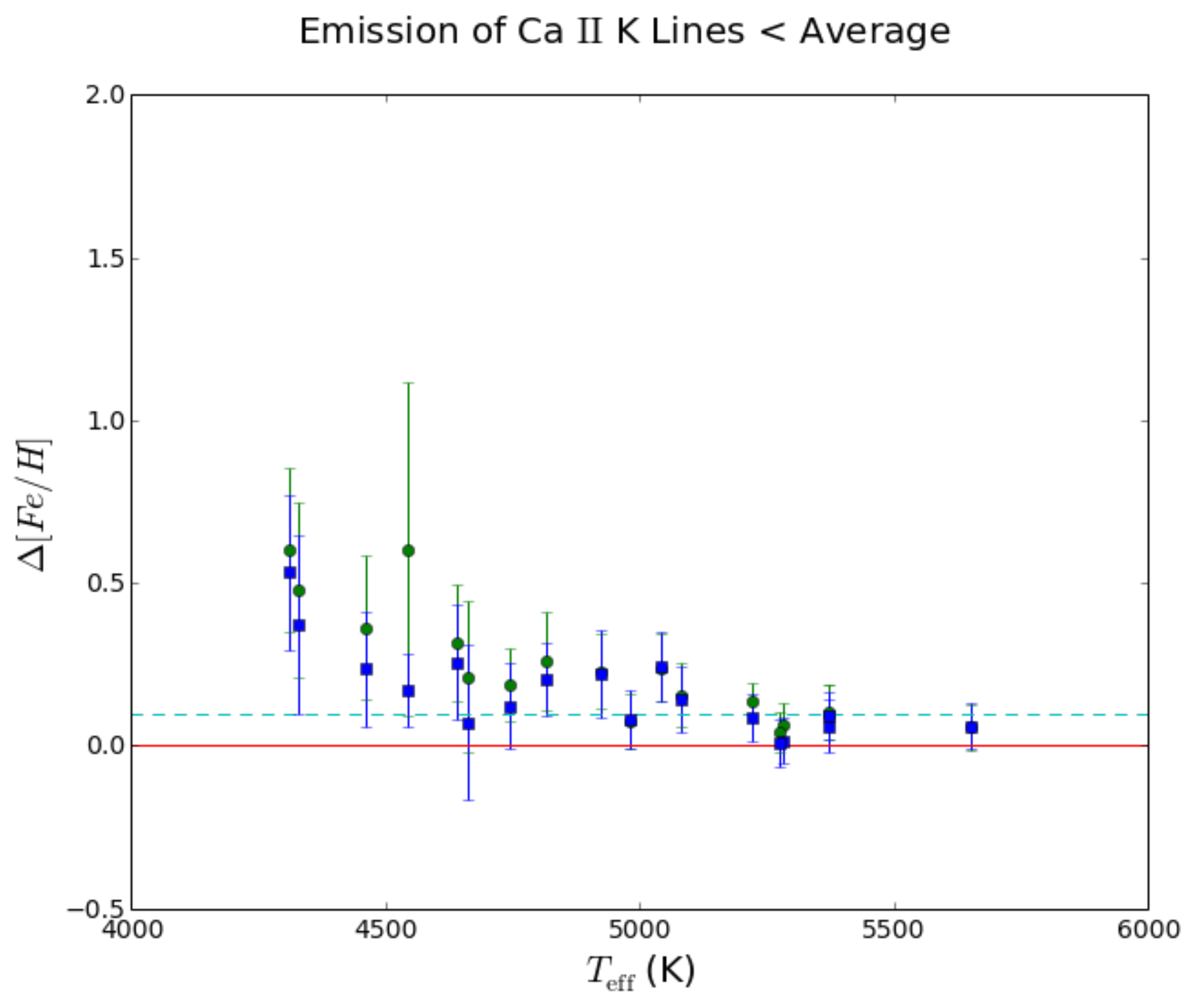}
    \caption{The difference in iron abundance (Fe {\scriptsize{II}} - Fe {\scriptsize{I}}) of our binary star secondaries where the emission as indicated by Ca {\scriptsize{II}} K 3933.6 {\AA} is \textit{less} than average. The green circles represent our initial measurements before the 6 Fe {\scriptsize{II}} bad lines are removed from our calculations. The blue squares are the results after the removal of these lines.}
    \label{CaKLess}
  \end{minipage}
  \hfill
  \begin{minipage}[b]{0.45\textwidth}
    \includegraphics[width=\textwidth]{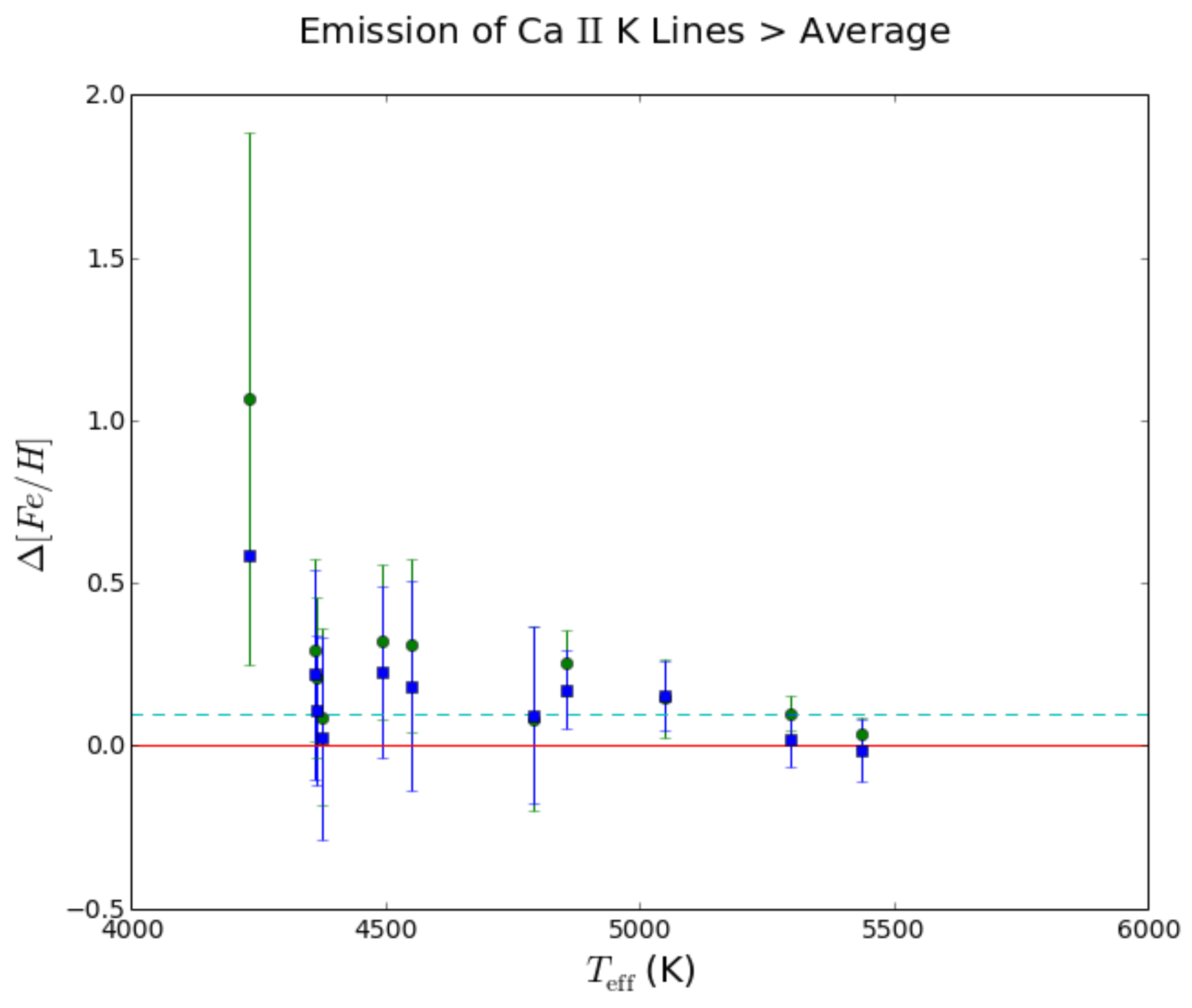}
    \caption{The difference in iron abundance (Fe {\scriptsize{II}} - Fe {\scriptsize{I}}) of our binary star secondaries where the emission as indicated by Ca {\scriptsize{II}} K 3933.6 {\AA} is \textit{greater} than average. The green circles represent our initial measurements before the 6 Fe {\scriptsize{II}} bad lines are removed from our calculations. The blue squares are the results after the removal of these lines.}
    \label{CaKGreater}
  \end{minipage}
\end{figure*}

Upon close inspection, there is no obvious or profound difference between stars with Ca {\scriptsize{II}} H+K lines above average and those below average; the only subtlety is that there are a few more stars that fall below average for both the H+K lines, therefore deeming them the ``older" stars relative to our sample. Comparing the figures, both the young and old exhibit the upward discrepancy trend towards lower $T_\mathrm{eff}$, and the rate of this increase is similar, ultimately indicating weak to no correlation between activity and iron abundance.

%%%%%%%%%%%%%%%%
%%%%%%%%%%%%%%%%

\subsubsection{Li Absorption}

\begin{figure*}[!tbp]
  \centering
  \begin{minipage}[b]{0.45\textwidth}
    \includegraphics[width=\textwidth]{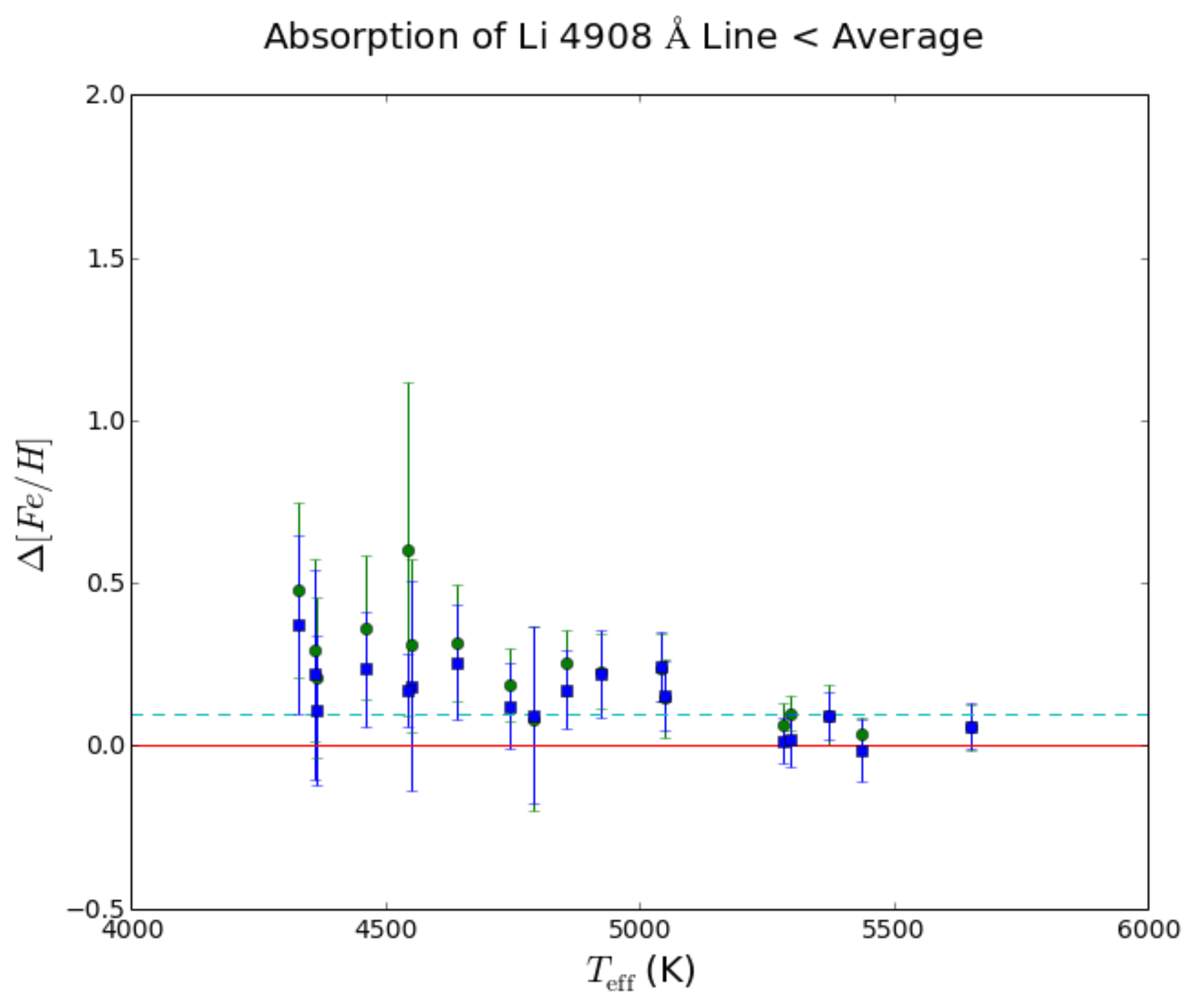}
    \caption{The difference in iron abundance (Fe {\scriptsize{II}} - Fe {\scriptsize{I}}) of our binary star secondaries where the emission as indicated by Li 4908 {\AA} is \textit{less} than average. The green circles represent our initial measurements before the 6 Fe {\scriptsize{II}} bad lines are removed from our calculations. The blue squares are the results after the removal of these lines.}
    \label{LiLess}
  \end{minipage}
  \hfill
  \begin{minipage}[b]{0.45\textwidth}
    \includegraphics[width=\textwidth]{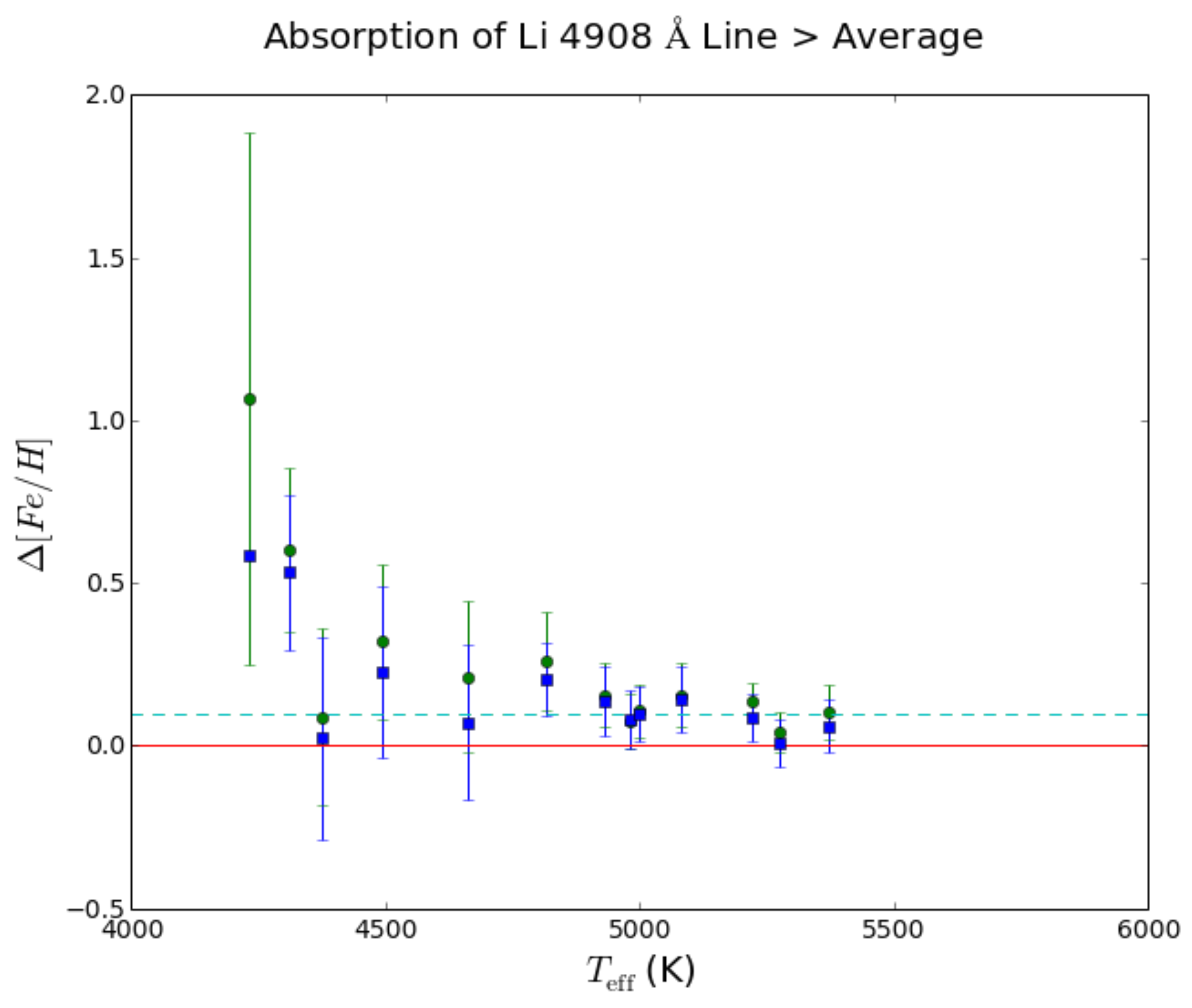}
    \caption{The difference in iron abundance (Fe {\scriptsize{II}} - Fe {\scriptsize{I}}) of our binary star secondaries where the emission as indicated by Li 4908 {\AA} is \textit{greater} than average. The green circles represent our initial measurements before the 6 Fe {\scriptsize{II}} bad lines are removed from our calculations. The blue squares are the results after the removal of these lines.}
    \label{LiGreater}
  \end{minipage}
\end{figure*}

The same process used to analyze the Ca {\scriptsize{II}} H+K lines was then applied to the Lithium absorption line at 6708 {\AA}. Again, we took a rough measurement of the depth of the absorption line for all secondaries, from which we calculated an average and separated our stars into two groups: ones which have less absorption than the average (presumed to be the older stars in the group) and ones which have more absorption than the average (presumed to be the younger stars in the group). Figures \ref{LiLess} and \ref{LiGreater} depict these two groups. 

As we observed in the Ca {\scriptsize{II}} H+K, there is no immediate or obvious relation which relates activity and our discrepancy. This is unsurprising, as typically the stars which had lower emission in Ca {\scriptsize{II}} H+K had lower Li absorption, indicating that these stars are, in fact, older in our sample. Likewise, the stars which had higher than average emission in Ca {\scriptsize{II}} H+K had higher than average Li absorption. If there is in fact a correlation or some connection linking the activity in the upper atmospheres of these stars and our perceived anomaly, it is not a strong one. Granted, our analysis of these age indicators was very approximate. A detailed measurement of EWs could possibly yield a more obvious correlation. However, it is doubtful that such a investigation would produce significantly different results, considering that we still see a discrepancy of up over 0.5 dex with stars around $T_\mathrm{eff}$ = 4300 K even after major line-blending considerations. Thus, we can move on to our final investigation: metallicity.

%%%%%%%%%%%%%%%%%%%%%
%%%%%%%%%%%%%%%%%%%%%

\subsection{Metallicity}
\begin{figure*}[!tbp]
  \centering
  \begin{minipage}[b]{0.45\textwidth}
    \includegraphics[width=\textwidth]{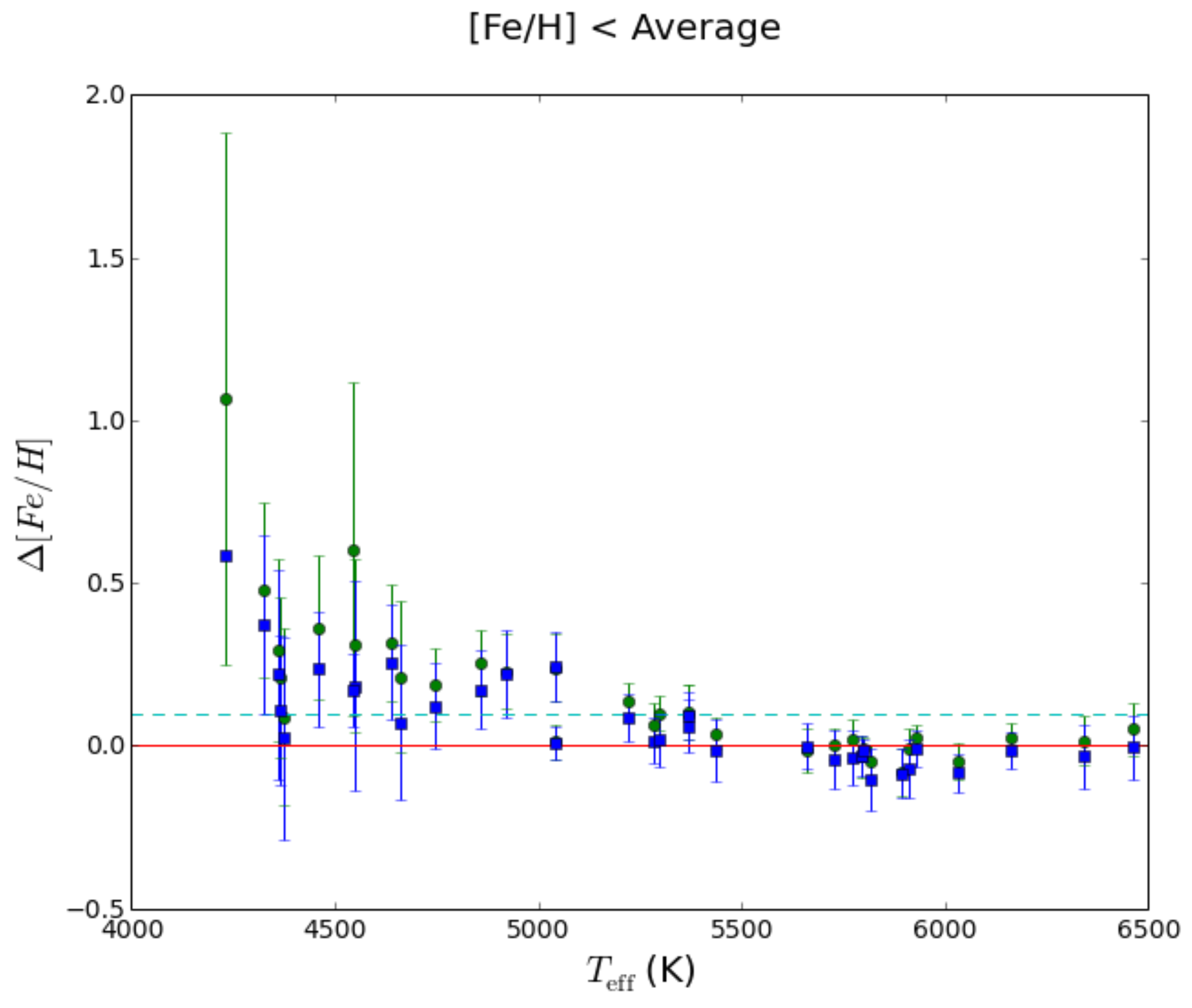}
    \caption{The difference in iron abundance (Fe {\scriptsize{II}} - Fe {\scriptsize{I}}) of our binary star sample where the [Fe/H] is \textit{less} than average when accounting for all stars in the sample. The green circles represent our initial measurements before the 6 Fe {\scriptsize{II}} bad lines are removed from our calculations. The blue squares are the results after the removal of these lines.}
    \label{[FeH]Poor}
  \end{minipage}
  \hfill
  \begin{minipage}[b]{0.45\textwidth}
    \includegraphics[width=\textwidth]{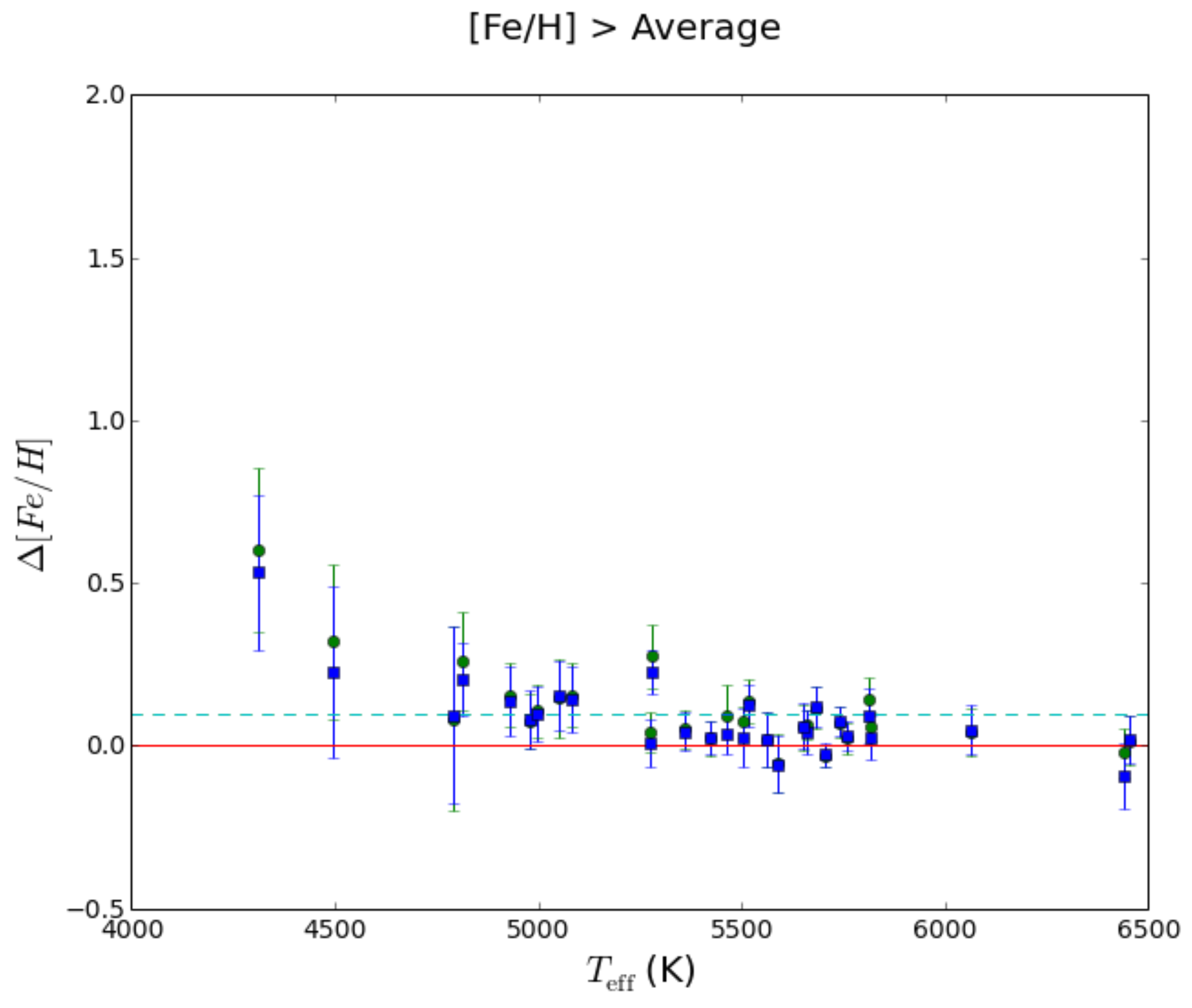}
    \caption{The difference in iron abundance (Fe {\scriptsize{II}} - Fe {\scriptsize{I}}) of our binary star sample where the [Fe/H] is \textit{greater} than average when accounting for all stars in the sample. The green circles represent our initial measurements before the 6 Fe {\scriptsize{II}} bad lines are removed from our calculations. The blue squares are the results after the removal of these lines.}
    \label{[FeH]Rich}
  \end{minipage}
\end{figure*}

Here, we compiled [Fe/H] for all 63 binary stars and found the average to be of sub-solar value: [Fe/H]$_{avg}$ = -0.0478. Subsequently, we divided all binary stars into two samples, the ``metal rich" sample of [Fe/H] $\geq$ -0.0478 and a ``metal poor" sample of [Fe/H] $\leq$ -0.0478. In both metallicity regimes, we plotted $\Delta$[Fe/H] vs. $T_\mathrm{eff}$, as shown in Figures \ref{[FeH]Poor} and \ref{[FeH]Rich}.

As was the case with both Ca {\scriptsize{II}} H+K emission and Li absorption, the morphology is qualitatively similar in both samples: not only do they exhibit the trend, but do so at approximately the same rate. Essentially, regardless of the metallicity of the star, as long as it is cool enough, the trend is present. This is consistent with the findings of \cite{schuler06:triplet}, when investigating [O/H] from the $\lambda = 7774$ {\AA} high-excitation triplet. They found that the Pleiades, which has [Fe/H] $\approx$ 0 according to \cite{Boesgaard90}, exhibits an even steeper trend than the Hyades, which has [Fe/H] = 0.20 according to \cite{Branch80}. From this, they predicted that the UMa moving group (which is also of sub-solar Fe value at [Fe/H] = -0.09 and an age coeval to that of the Hyades) would have an even steeper trend, assuming metallicity was a factor. But in fact, they discovered the opposite: the UMa moving group exhibits a trend shallower than that of the Pleiades. Additionally, the rate of the trend in [O/H] (not $\Delta$[O/H]) of the UMa moving group matches that of the Hyades. \par

For our sample, the Hyades are found to be of super-solar Fe metallicity of [Fe/H] = 0.098, and exhibit a slightly shallower trend than that of our binaries. At first glance, this small deviation could be indicative of metallicity. Under further inspection, we find that our ``metal rich" binary sample has an average metallicity of [Fe/H] = 0.093, making it comparable to the Hyades. However, it has a slightly steeper trend. If metallicity was a factor, this would not be the case. Therefore, both the study from \cite{schuler06:triplet} and our study leads us to believe that, albeit not conclusive, metallicity is likely not a contributing factor.

%%%%%%%%%%%%%%%%%%%%%
%%%%%%%%%%%%%%%%%%%%% 

\section{Concluding Remarks}
As was believed by the stellar abundance community, we found that line-blending was indeed the main cause for the discrepancy of iron abundance as inferred from Fe {\scriptsize{I, II}} lines. In our binary sample, it cut $\Delta$[Fe/H] by at most $\thicksim$ 0.5 dex, and by $\thicksim$ 0.6 dex in our Hyades sample. Though, we still find an observed discrepancy in both samples in stars with $T_\mathrm{eff}$ $\leq$ 4700 K, with a difference of up to $\thicksim$ 0.6 dex in the field binaries and $\thicksim$ 0.3 dex in the Hyades. The reason for this is still unknown. NLTE effects are far more prominent in warmer, solar-type stars where we do not observe the anomaly, whereas in the cooler stars these effects are, to our knowledge, either negligible or small but will not contribute to the remaining trend. Additionally, we do not find anything of interest arising from our study of Ca {\scriptsize{II}} H+K emission and Li absorption. Assuming younger stars in general would have stronger emission and absorption (which we admit is not a tight correlation), we grouped our samples into ``young" and ``old", and found the trend in both denominations, each having similar slopes. It must be noted that there is always the possibility that general limitations in our understanding of K-dwarf atmospheres could play a prominent factor, especially in regard to more in-depth NLTE line formation and atmospheric modeling. Advancements in these fields are awaited with interest.\par

Although this paper focused on Fe {\scriptsize{I}} and Fe {\scriptsize{II}} lines, \cite{schuler06:triplet} found the same discrepancy in [O/H] from the O {\scriptsize{I}} triplet at $\lambda$ = 7770 {\AA} in the Pleiades and Hyades. They employed multi-component 1D-LTE models, showing that starspots are a plausible source. They modeled stars with hot areas (corresponding to faculae/plages), quiescent areas, and cool areas (corresponding to starspots). The O {\scriptsize{I}} triplet lines are stronger at higher temperature, and so the hot component is required to reproduce the O {\scriptsize{I}} triplet observations as they are stronger than predicted by non-spot models. This ultimately leads to the enhanced oxygen abundances. Since these lines are insensitive to the cool areas, they are not useful to determine their properties such as surface coverage and temperature. In this sense, these starspot models are not fully constrained and therefore they have not yet been validated. \par

To address this deficiency, we propose future studies to observe K-dwarfs of the Hyades and Pleiades, specifically utilizing the first-overtone vib-rot OH lines at 1.555 $\mu$m, which are detectable in stars of the same temperature range as this paper. Further, the OH features are stronger at lower temperatures than the O {\scriptsize{I}} triplet, which makes them the ideal candidate to probe the cool starspots. Additional constraints could be provided by the wavelength dependency of the continuum flux emitted by the cool and hot areas. These observations will help determine starspot properties as a function of effective temperature and age. Together with optical spectra to establish a self-consistent framework, these observations will allow future studies to evaluate model uncertainties and devise survey strategies to best exploit the observations of K-dwarfs. \par

\section{Acknowledgements}
We would like to thank the John W. Cox Endowment for the Advanced Studies in Astronomy, as well as the McDonald Observatory/Department of Astronomy Board of Visitors Fund. Additionally, we are indebted to Dr. Lyudmila Mashonkina for private communication regarding NLTE effects and to Dr. Mike Montgomery for his help in an advisory role. 

\bibliography{Abundance_ref}

\end{document}